\documentclass[numbered]{trbunofficial}
\usepackage{graphicx}
\usepackage{subfig}
\usepackage{multirow}
\usepackage{url}
\usepackage{float}

\usepackage[hidelinks]{hyperref}

\AuthorHeaders{Bian, Zuo, Gao, Ozbay and Maggio}
\title{A-TEAM: Advanced Traffic Event Analysis and Management Platform for Transportation Data-Driven Problem Solving}

\author{%
  \textbf{Zilin Bian, Ph.D., Corresponding Author}\\
  Graduate Research Assistant, C2SMART Center,\\
  Department of Civil and Urban Engineering,\\
  Tandon School of Engineering, New York University, \\
  6 MetroTech Center, 4th Floor, Brooklyn, NY, 11201, USA\\
  \hfill\break
 \textbf{Dachuan Zuo, M.S.}\\
  Graduate Research Assistant, C2SMART Center,\\
  Department of Civil and Urban Engineering,\\
  Tandon School of Engineering, New York University, \\
  6 MetroTech Center, 4th Floor, Brooklyn, NY, 11201, USA\\
  \hfill\break
  \textbf{Jingqin Gao, Ph.D.}\\
  Postdoctoral Associate, C2SMART Center,\\
  Department of Civil and Urban Engineering,\\
  Tandon School of Engineering, New York University, \\
  6 MetroTech Center, 4th Floor, Brooklyn, NY, 11201, USA\\  
  \hfill\break
  \textbf{Kaan Ozbay, Ph.D.}\\
  Professor \& Director, C2SMART Center,\\
  Department of Civil and Urban Engineering,\\
  Tandon School of Engineering, New York University, \\
  6 MetroTech Center, 4th Floor, Brooklyn, NY, 11201, USA\\ 
  \hfill\break
  \textbf{Matthew D. Maggio, B.S.}\\
  System Engineer, C2SMART Center,\\
  Department of Civil and Urban Engineering,\\
  Tandon School of Engineering, New York University, \\
  6 MetroTech Center, 4th Floor, Brooklyn, NY, 11201, USA\\
  \hfill\break
}

\begin{document}
\maketitle

\section{Abstract}
The rapid growth in terms of the availability of transportation data provides great potential for the introduction of emerging data-driven methodologies into transportation-related research and development efforts. However, advanced data-driven models, such as artificial-intelligence based approaches, usually contain complicated modeling structures and require strict data formats along with a very complex execution environment.  It is thus often challenging to deploy and implement such data-driven models in a real-world environment. Moreover, a full-fledged application requires not only well developed and calibrated models, but also efficient connections with back end infrastructure such as large databases and front end utilities, such as a user-interface. This paper introduces a novel platform which provides an integrated architecture for deploying multi-purpose real-time traffic management applications. Inspired by the concept of modular design in software system development, this paper proposes a modular platform allowing users to customize their mission specific needs and preferences. The developed platform is capable of incorporating flexible user-provided models and/or data with the ultimate goal of deploying them as a complete application ready for real-world use. To illustrate this novel modular software system concept, this paper presents a work zone management and coordination application that is built upon the developed implementation platform to provide useful decision support to traffic engineers.

\noindent\textit{Keywords}: Modular design, Spatiotemporal analysis, Traffic event, Coordination analysis, Transportation management tool
\newpage

\section{Introduction}

There has been a significant growth of both infrastructure based and mobile data collection technologies. As a result, massive amounts of transportation data have become available, making it possible to improve and enhance a variety of transportation applications used for on and off line decision making. As part of this positive trend in terms of the availability of big transportation data, emerging data-driven models based on artificial intelligence (AI) and machine learning (ML) have been increasingly adopted by the transportation engineering community to deal with various important transportation problems, such as real-time forecasting of traffic state \cite{ma2015large}, detecting time-dependent changes in terms of traffic mobility \cite{gao2022new, bian2021time}, predicting duration and other characteristics of traffic accidents \cite{fu2019titan}, and so on. 

A complete transportation application that can be used as an integrated tool to provide decision support is usually designed using a conventional three-tier software design architecture called Model-View-Controller (MVC) \cite{krasner1988description}. The MVC requires the identification of interactions between application modules such as databases, model deployment and implementation of web-based protocols. Despite the success of data-driven models in addressing challenging transportation problems at the research end, the deployment of these data-driven models in functioning software is still limited by the lack of some key software modules, as described in the context of MVC design architecture. Therefore, implementing data-driven models as part of deployable applications still faces major challenges. 

These challenges usually include 1) the development of a user interface, 2) the conversion from offline to an online model, 3) the connection to databases for continuous data collection, storage, and processing, 4) the capability to adapt to new development in the underlying data models or environment (e.g., a new open source software library), and 5) the scalability and flexibility to incorporate new functionalities driven by changing user needs and preferences. For example, the processing of the transportation data as part of research is well-understood. In many research prototypes, the data collection and processing efforts are conducted offline and only the necessary portion of processed data is needed for modeling, training, and testing stages. However, the deployment of a real-world transportation application usually requires the data acquisition and processing tasks which feeds the application to be processed and aggregated in an automated manner that works in real-time. Another example is that to operationalize a data-driven model, commonly used back-end languages for server scripting (e.g., PHP) may not be directly compatible with other non-graphical languages (e.g., Python builds). It can also be extremely time-consuming to implement specific data-driven models at the application end given their complex model structure and strict requirements with respect to the computing environment and required data format. 

To cope with the above practical, yet important challenges, we propose a highly customizable and adaptive platform referred to as Advanced Traffic Event Analysis and Management (A-TEAM) platform. The A-TEAM platform is a framework specifically designed for the transportation domain that aims to accelerate the path from research prototyping to production deployment. The goal of the A-TEAM platform is to reduce the efforts required from both traffic researchers and practitioners for developing traffic event-related applications while allowing them to employ the power of various types of AI/ML based  data-driven models. The main objectives of the A-TEAM platform can be summarized as follows:

\begin{itemize}
    \item Develop a modular system allowing users to customize their transportation software application according to their needs and preferences. This modular system contains functional modules with multiple levels of customization, enabling the scalability and readiness to incorporate novel data-driven technologies. 
    \item Design a spatiotemporal data mining module which contains data acquisition, data processing and data fusion functionalities, support standardized spatiotemporal data to help researchers and practitioners in model development, training, and on-line data management. 
    \item Design an extendable modeling scheme which supports multiple computational libraries and a flexible input-output channel between the front end and the back end of the web application.
    \item Design a friendly web user-interface to allow users to perform multiple input and output tasks based on their customized user needs.
    \item Illustrate the feasibility of actually incorporating the above objectives in a working product through a finished traffic application designed to facilitate the task of traffic management based on the proposed A-TEAM platform.
\end{itemize}

\section{Literature Review}\label{sc:literature}
There has been a number of visual platforms or software tools used in the transportation world for the analysis \cite{ozbay2013assist, zhu2021save} and management \cite{bartin2017work} of traffic-related events. QuickZone \cite{quickzone}, WISE \cite{WISE}, CIA\cite{CIA}, and GLRTOC\cite{GLRTOC} are examples of visual analytics tools for work zone analysis. TraCS\cite{TraCS} and TIM-BC\cite{ma2016user} are examples of such tools designed for accident and incident management. These kinds of software provide graphical toolkits that allow practitioners to conveniently conduct spatial and temporal analysis and are capable of carrying out mathematical models to evaluate the impact and corresponding response plans. However, most traditional software including \cite{quickzone} and \cite{CIA} only supports the use of parametric models, such as deterministic queuing models \cite{jiang2003freeway, chitturi2008methodology}, which requires comprehensive data inputs such as traffic volume, road geometry, and details of the traffic event. However, these parametric models suffered in the issues of over-estimation \cite{jiang2003freeway} and limited scope when adapting to different modeling purposes \cite{ramezani2011methodology}. For example, adapting these tools to different tasks can be challenging since they are usually designed for a predetermined usage such as traffic analysis or work zone management. Furthermore, the traditional deterministic queuing models rely on comprehensive traffic volume information, which is not easy to acquire through typical data collection efforts. \newline

Data fusion is one of the solutions that can relieve the users from the cumbersome work of manually acquiring  and modifying data and its parameters. This technique fuses information from multiple inter-related datasets, which allows various data to be combined to improve the data enrichment of individual databases and aid with the analysis by providing more diverse data features. A handful of recently developed systems have utilized this technique including \cite{dickerson2016work}, \cite{wen2018integrated}, \cite{perez2019developing}, \cite{wongsuphasawat2009visual}, and \cite{weissman2008plan4safety}. The web-based work zone analysis tools in \cite{dickerson2016work} and \cite{wongsuphasawat2009visual} and the crash decision-support tool in \cite{weissman2008plan4safety} integrate information about the traffic, the work zone itself, and the roadway to fulfill the designed functionalities, such as work zone status tracking and the evaluation of a response plan. The incident management system in \cite{wen2018integrated} has a module to combine and cleanse traffic data from multiple sources. The web-based emergency management system in \cite{weissman2008plan4safety} fuses data from both social and economical categories to analyze their correlations with the incident. \newline

Furthermore, data fusion techniques improve data diversity and make feasible the use of data-driven models in various analysis tasks. A number of recent incident analysis tools in the literature have attached great importance to machine learning models and the functionalities. \cite{kalamaras2017interactive} proposes a prototype traffic analysis platform which implements machine learning models pre-trained using PeMS data \cite{pems} to predict the traffic pattern and facilitate the evaluation of management strategies. Their implemented models include k-nearest neighborhood (k-NN)\cite{myung2011travel} and support vector regression (SVR)\cite{wu2004travel}. Another example is the incident management platform developed by \cite{wen2018integrated} which features a predictive module to analyze incident impacts using Dynamic Bayesian Network (DBN)\cite{ghahramani1997learning}. The prediction output is further utilized simultaneously with a simulation module to evaluate the response plan. Moreover, \cite{wang2019artificial} proposes a prototyped AI platform to solve traffic prediction problems for research purposes. They incorporated pre-trained neural network models such as Long-short term memory (LSTM)\cite{hochreiter1997long} and gradient recurrent unit (GRU)\cite{chung2014empirical} to supply as prediction modeling for traffic forecasting tasks. \newline

However, all of the above reviewed systems can barely bridge the gap between the demand of researchers and practitioners because they lack some, or in some cases all, of the functionalities cited below. First, an ideal system should be flexible to allow users to customize their own datasets and to be scalable to accommodate new  models as they become available according to their most current needs. Second, it should also be able to retrieve real-time updates of event and traffic data while having the capability of solving real-world transportation problems that require a timely solution. The prototype of \cite{wang2019artificial} satisfies the first requirement since it allows users to upload and validate their own deep learning models for traffic prediction with their choice of datasets. However, their tool is mainly research oriented and thus not capable of using data in real-time. It also lacks an interactive user interface which prevents user-friendly visualizations when performing task based analysis like other reviewed tools. Other systems including  \cite{wongsuphasawat2009visual} and \cite{dickerson2016work} can technically accept customized datasets since they allow data to be uploaded off-line, however this requires extensive back and forth between researchers and the software developers to ensure successful use of these new datasets.    

\section{Framework of the A-TEAM Platform}\label{sc:framework}

\begin{figure}[H]
    \centering
    \includegraphics[width=1\textwidth]{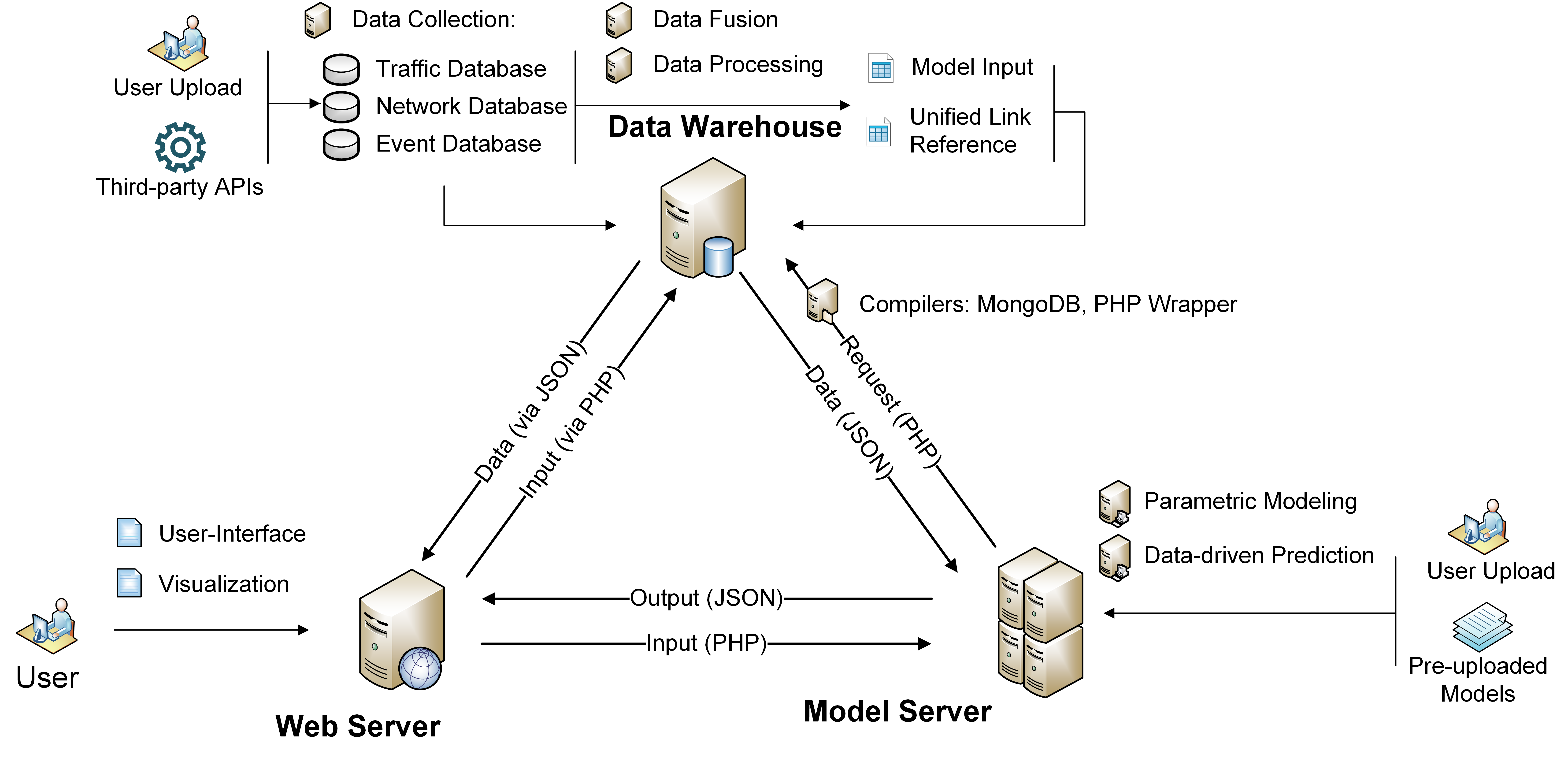}
    \caption{Designed architecture of A-TEAM platform}
    \label{fig:D-TEAM-framework-workflow}
\end{figure}

The A-TEAM platform is designed to bridge the gap between the research prototyping and the application deployment. The objective of A-TEAM platform is to help traffic researchers and practitioners to deploy their theoretical solutions to various cutting-edge transportation problems in an applicable and operational system. The architecture of A-TEAM platform (Fig.\ref{fig:D-TEAM-framework-workflow}) consists of three key components: 1) Data warehouse, 2) Web server and 3) Model server. The data warehouse is designed with the purpose of maintaining and managing the corresponding datasets. It contains modules as data collection, data processing and data fusion.  The web server is designed to allow users to input, interact and collect outputs in the front end. Modules in the web server include visualization and user-interface. The model server provides the computational power by sending the modeling results from the back end (server) to the front end (browser). The model server provides a compatible computing environment using high performance computation resources, supporting users to run both parametric modeling and data-driven modeling approaches.

\subsection{Modular Design}
The A-TEAM is designed using the concept of " Multi-data, Multi-model, Multi-task and Customization (MMM\&C). "C" represents the customization, which is inspired by the proposed modular design. The modular design is an approach used to design a comprehensive and complex system/application, such as an automobile, a smart phone or computer software. Such systems/applications with the modular design are usually disassembled into several individual and separate sub-modules. These sub-modules are integrated using unified standards and can be reassembled and reused to produce different products without the effort of re-designing the whole system from scratch. Analogous to building blocks, with appropriate combinations and organizations, these sub-modules are flexible and easy to be customized according to user needs or preferences. 

Therefore, we designed and implemented a number of highly customizable modules with different levels of integration efforts. To be specific, the level of integration indicates the difficulties of customizing the software modules and a higher level of customization (e.g., level 3) can support more customized needs. 

\begin{figure}[H]
    \centering
    \includegraphics[width=1\textwidth]{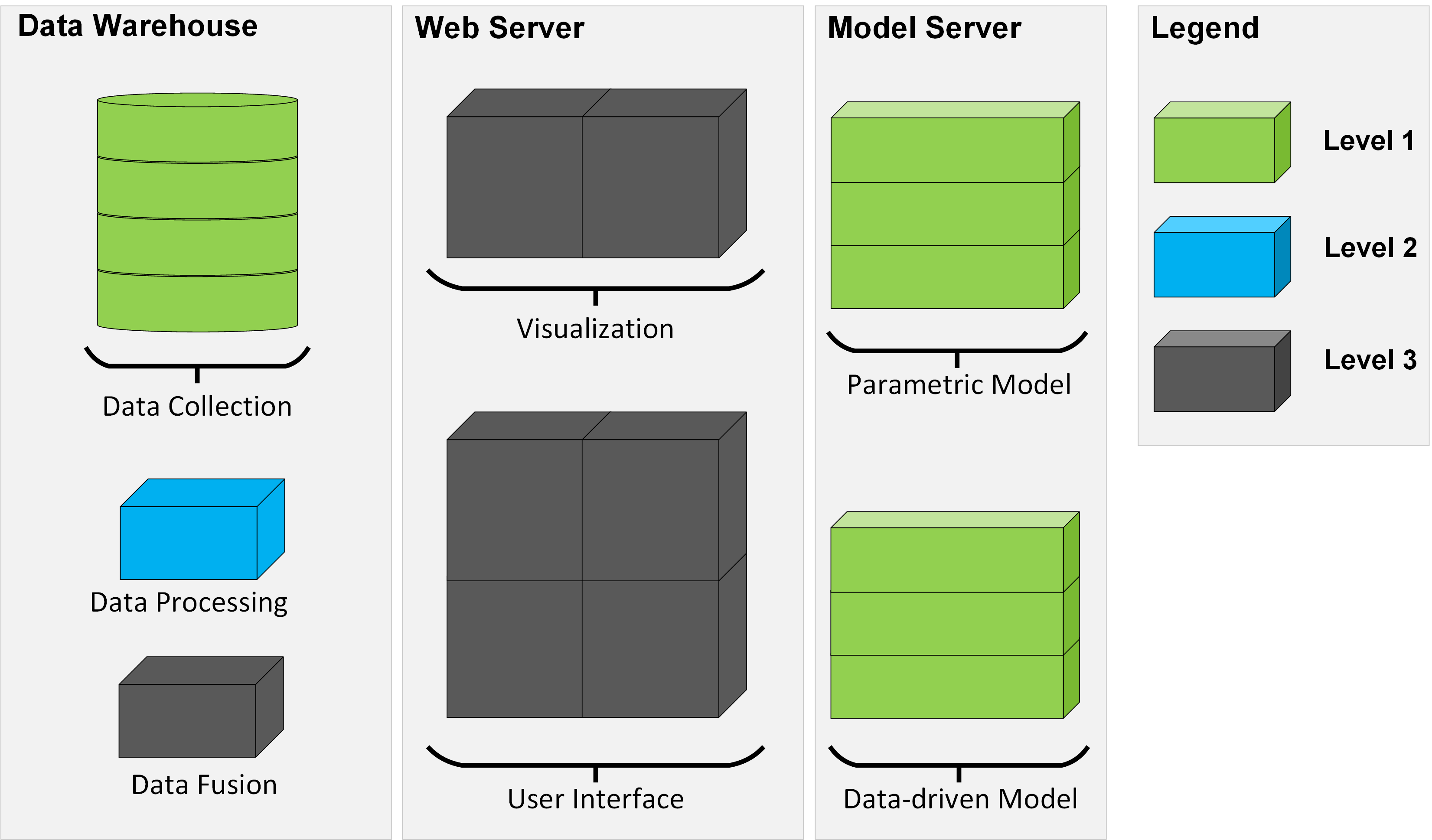}
    \caption{Designed concept of A-TEAM platform}
    \label{fig:A-TEAM-concept}
\end{figure}

As shown in Fig.\ref{fig:A-TEAM-concept}, the modules are demonstrated in terms of  square and rectangular shapes with three different colors. Level 1 (in green) represents the easiest level of integration, indicating users do not need to put significant effort into customizing and changing the modules. At Level 1, users only need to provide inputs such as data and model files, then A-TEAM will incorporate these instantly. Level 2 (in blue) indicates an intermediate effort of integration in customizing such modules. At this level, users need to work with the semi-product provided by A-TEAM or make changes intermediately in order to operationalize the module. For example, A-TEAM has a data processing sub-module, offering pre-processed data which can assist users in training their own models. Level 3 (in grey) represents the highest level of interference. At this level, A-TEAM provides a complete package of functional modules, requiring the most significant software development effort if users want to customize the modules to cope with their specific user needs or preferences. This can include efforts in modifying both front end and back end aspects of the A-TEAM platform. For example, if users would like to add a new analysis function to A-TEAM, which requires new combinations of various models and data, they may not only work with model and database at the back end but also work with the pre-designed user interface at the front end. Such in-depth customization requires significant effort, asking users to closely interact with A-TEAM developers.


\subsection{Data warehouse}

The data warehouse is designed to fulfill two functionalities. First, it collects and maintain various types of data regarding incidents, roads, and traffic conditions which are retrieved from multiple sources. Second, it processes and prepares the data for the needs of visualization and modeling. These functionalities are fulfilled by three modules -- data collection, data fusion, and data processing. The data warehouse is implemented with MySQL which maintains offline data, PostgreSQL which stores geographical information, and MongoDB which communicates with the model server. The data warehouse also connects external APIs to maintain online data. 

\subsubsection{Data Collection}

Three distinct databases related to traffic event analysis and management are contained in the data warehouse: traffic database, infrastructure / network database, and event database.\newline
\begin{itemize}
\item \emph{Traffic database} contains information about traffic conditions including traffic volume and travel speed. The traffic data is collected from NPMRDS dataset \cite{sunyavail} in the current implementation of A-TEAM. The travel speed data is collected by aggregating the travel speed of individual probe vehicles by a time interval of 5 minutes, 15 minutes, and 1 hour. \newline
\item \emph{Infrastructure / Network database} stores geographical information of the roadways including roadway geometry, referencing ID, and characteristics such as number of lanes and lane width. This type of spatial data is collected from NPMRDS dataset and GIS data clearing house of the New York State Department of Transportation (NYSDOT) \cite{gisnysdot} in the current implementation of A-TEAM. \newline
\item \emph{Event database} acquires and stores information regarding various types of incidents such as work zones, accidents, or traffic anomalies from real-time feeds (e.g., 511NY) \cite{511ny} and off-line tabular data files provided by the New York City Department of Transportation (NYCDOT). The information includes the type, location, and duration of events. \newline 
\end{itemize}

These databases are designed to be highly customizable and friendly to both practitioners and researchers. Fig.\ref{fig:A-TEAM-concept} labels the data collection process with a level 1 customization level because the users are allowed to upload their own data files or use external APIs. Further, the databases support traffic and event data to be added and updated in  real-time, which is essential for the development of an incident management tool. Additionally, researchers are able to use their own sources of traffic and event data to further utilize more advanced functionalities provided by A-TEAM, as long as the data satisfies the minimum field compatibility requirements in the various tabular formats such as Microsoft Excel (.xlsx), text document (.txt) and JavaScript object notation (.json): the  required fields for event data include event type, coordinates, and duration, and the minimum required fields for traffic data include speed, volume, and the geometric and temporal features of the corresponding roadway.

\subsubsection{Data Fusions}

The data warehouse module involves two data fusion methods to prepare a unified data format for visualization, analysis and modeling input purposes: 1) event-road matching method and 2) traffic-road matching method. These data fusion methods are implemented in advance to be part of A-TEAM and to work in an automated manner. When new data becomes available, the data fusion module automatically tracks and fuses the data using a unified road way referencing system that is based on the indexing of the spatial database of A-TEAM. The details of data fusion are as follows. 

The first method maps event data to the roadway in the infrastructure / network database and creates a unified event-road table consisting of a unique roadway ID and the attributes of both the event and roadway. Since the raw data of events is stored as points, this method involves a step to find the most likely roadway from the infrastructure / network database where the event has occurred. The software evaluates the likelihoods of an event-road match by checking the name and type of the roadway as well as the street information contained in the event data. 

The second method maps the traffic data to the roadways and creates a unified traffic-road table with a unique roadway ID and includes the information of both the traffic and the roadway. The traffic data and the spatial data that are imported from different sources tend to use different indices for their roads, requiring such discrepancies to be carefully handled. A-TEAM finds the best match between the roadway and the traffic data by running a similarity check on the street information contained in the traffic database and the characteristics of roadways in the spatial network database. Information such as road name and road type is compared to complete the matching process.

With the use of the above two data fusion methods, the event data, traffic data and infrastructure / network data are cross-indexed and can be referenced using an identical index. For example, when a traffic event is selected by users, the roadway information where the event is located and the traffic information during the occurrence of the event become available. When a roadway link is selected, the roadway information and the historical events that occurred on this link also become available to users. It is worth noting that the customization level of data fusion module is level 3, which means that modifying the methods of data fusion is possible but requires the user to collaborate with the A-TEAM developer.

\subsubsection{Data Processing}

The data processing module covers all of the steps that are needed to bridge the gap between the output of data fusion and the input of visualization and models. This module is determined as Level 2 customizable in Fig.\ref{fig:A-TEAM-concept} since the inclusion of data processing steps depends on the nature of data and the needs of functions to be performed. 

\begin{figure}[H]
    \centering
    \includegraphics[width=0.8\textwidth]{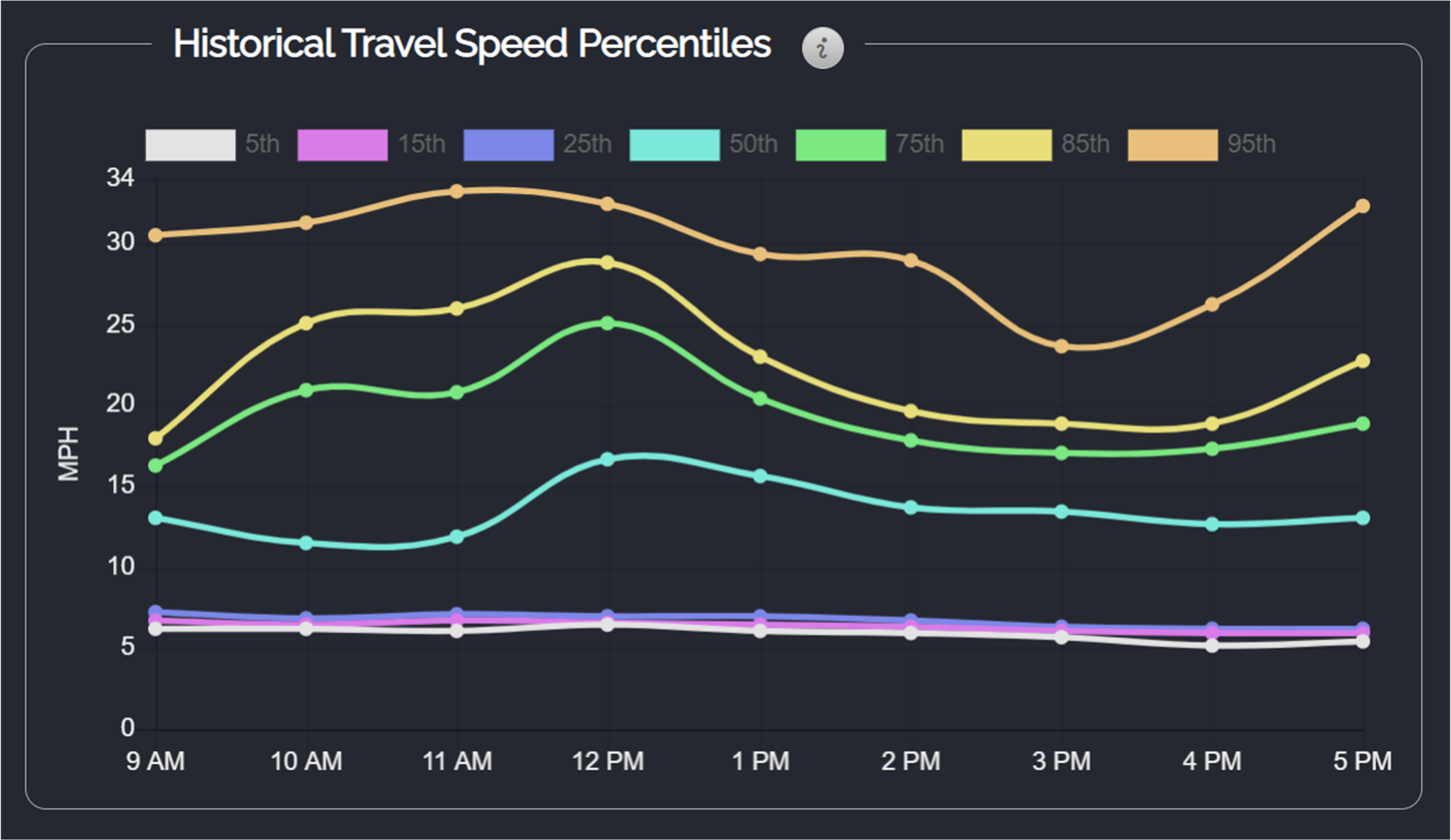}
    \caption{Example of historical travel speed percentiles}
    \label{fig:A-TEAM-SPEED-PERCENTILE}
\end{figure}

Two data processing actions have been implemented in A-TEAM. When a planned or unplanned incident is being dealt with, the first action is automatically carried out to calculate the percentile of travel speed distribution on the affected roadway over a time horizon that is determined by the user. Travel speed percentiles are useful metrics indicating the historical performance of a selected roadway. In most cases, the 5th percentile speed can represent the worst situations in the history when the roadway is extremely congested, the 85th percentile speed represents the prevalent situations in the history, and the 95th percentile represents the most ideal situation when the traffic is low. Fig.\ref{fig:A-TEAM-SPEED-PERCENTILE} shows the change of the historical speed percentiles of a selected roadway over 9 hours.

The second action is initiated when the user is trying to download data from a pre-selected dataset. This action retrieves all requested data from existing databases, performs a quality check by clearing off missing fields and errors, and integrates the cleansed data to a unified format. This function is designed to respond to the needs of researchers who will use A-TEAM data to pre-train or validate their own model in the offline mode.

\subsection{Model server}

The model server provides two categories of models to estimate or predict the impact of a planned or unplanned incident on the local traffic. The first category includes various parametric models, which estimate incident induced delays based on predetermined parameters. The second category encompasses data-driven models, which are mainly AI/ML models that are pre-trained and validated off-line using historical data before being deployed in the model server. 

This module is considered Level 1 customizable (requires the least effort from users) since the A-TEAM platform provides an interface for the users to upload their own pre-trained model files. A-TEAM installs high performance computation infrastructure including clusters of central processing unit (CPU) and graphical processing unit (GPU), and supports the execution of model files across various programming languages such as JavaScript and Python. More importantly, A-TEAM incorporates cross-language wrappers allowing connections between different types of programming languages. For example, a PHP-Python wrapper is installed to support communication between PHP and Python, MongoDB has the extensions enabling the conversation between JSON and Python. This allows data-driven model programming in the back end (usually built in Python) to take input features stored as PHP/JSON from the front end web browser or data warehouse and send estimated outputs back to the front end.
The programming environment and supported libraries incorporated in the model server are shown in Table.\ref{tab:model}. There are four parametric models \cite{bartin2012interactive,jiang2003freeway,chitturi2008methodology,ramezani2011methodology} currently implemented in the model server for the purposes of estimating the traffic impacts caused by the traffic events. In addition, there are three AI/ML models implemented or validated in the model server, including XGBoost\cite{chen2016xgboost}, Random Forests\cite{breiman2001random}, and Artificial Neural Networks (ANN)\cite{wang2003artificial}. An explanation of the outputs, environments, and required packages of these models are provided in Table \ref{tab:model}.

\begin{table}[H]
\caption{Explanation of all implemented models in A-TEAM}
\centering
\begin{tabular}{|l|l|l|l|}
\hline
Model          & Ouput                                                                                 & Environment                                                                                          & Library                                                                                \\ \hline
Parametric1\cite{bartin2012interactive}      & \multirow{4}{*}{\begin{tabular}[c]{@{}l@{}}Total Delay\\ Queue Length\end{tabular}}   & \multirow{4}{*}{\begin{tabular}[c]{@{}l@{}}PHP\\ JSON\end{tabular}}                                  & \multirow{4}{*}{\textbackslash{}}                                                      \\ \cline{1-1}
Parametric2\cite{jiang2003freeway}      &                                                                                       &                                                                                                      &                                                                                        \\ \cline{1-1}
Parametric3\cite{chitturi2008methodology}      &                                                                                       &                                                                                                      &                                                                                        \\ \cline{1-1}
Parametric4\cite{ramezani2011methodology}      &                                                                                       &                                                                                                      &                                                                                        \\ \hline
XGBoost\cite{chen2016xgboost}        & \multirow{3}{*}{\begin{tabular}[c]{@{}l@{}}Travel Speed\\ Congestion level\end{tabular}} & \multirow{3}{*}{\begin{tabular}[c]{@{}l@{}}Python 3.6.1\\ MongoDB\\ Python-PHP Wrapper\end{tabular}} & \multicolumn{1}{r|}{\multirow{2}{*}{Scikit-learn 1.1.1}}                               \\ \cline{1-1}
Random Forests\cite{breiman2001random} &                                                                                       &                                                                                                      & \multicolumn{1}{r|}{}                                                                  \\ \cline{1-1} \cline{4-4} 
ANN\cite{wang2003artificial}            &                                                                                       &                                                                                                      & \multicolumn{1}{r|}{\begin{tabular}[c]{@{}r@{}}PyTorch 1.5.0\\ Cuda 10.0\end{tabular}} \\ \hline
\end{tabular}

\label{tab:model}
\end{table}

\subsection{Web server}

As mentioned above, the design of the web server follows the conventional MVC approach. The web server fulfills the roles of viewer and controller, as it supports the visualization of data and computational results. More specifically, it provides a graphical user-interface including an interactive and user-friendly visualization tool, along with a number of analysis tools to manipulate the data and run the models. The physical design of A-TEAM enables the communications between the web browser and the web server via AJAX calls, which relieves the user from constantly refreshing the browser when they initiate a complex computation task.

\subsubsection {User-interface} 

\begin{figure}[t]
    \centering
    \includegraphics[width=0.8\textwidth]{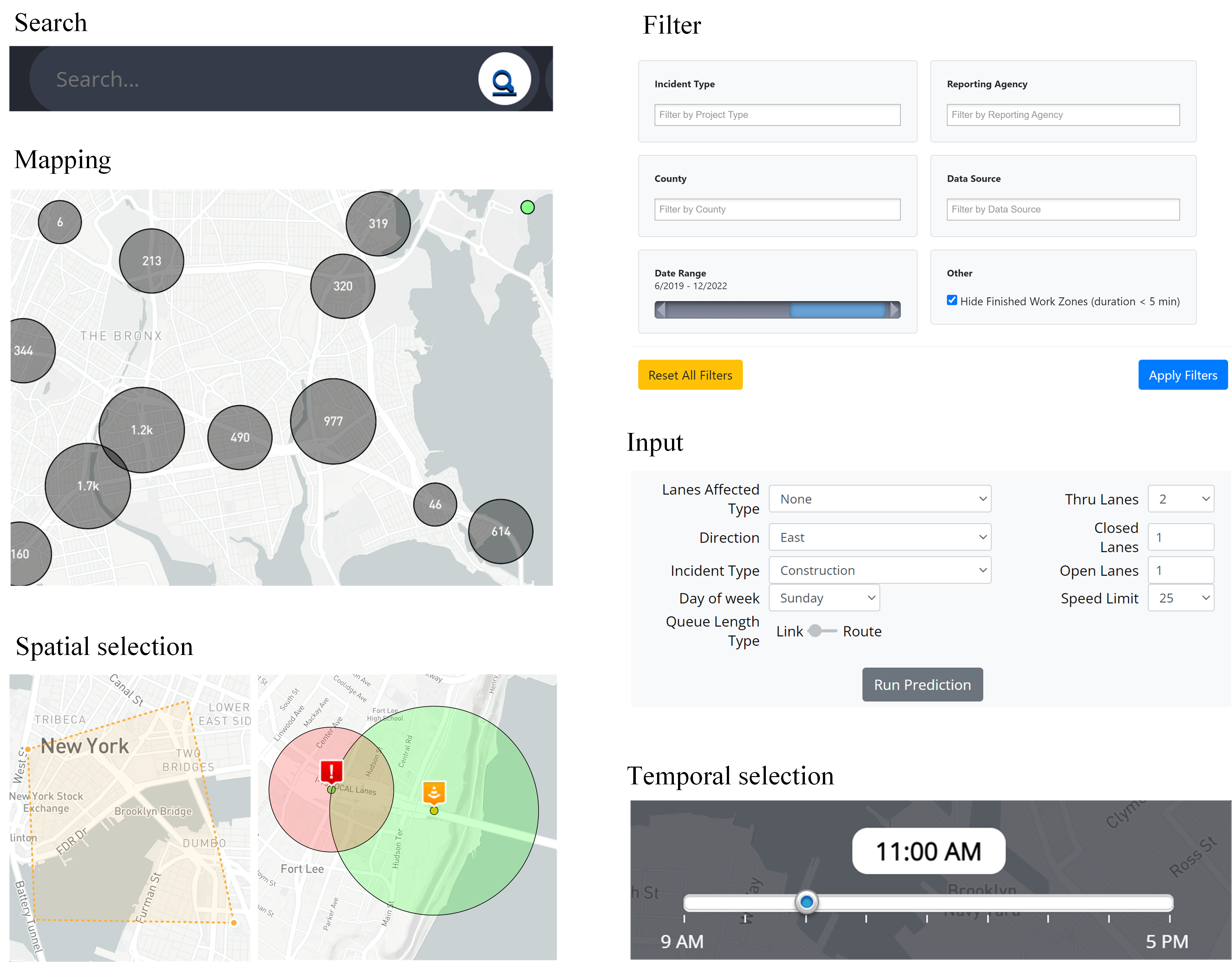}
    \caption{Examples of all UI functionalities}
    \label{fig:A-TEAM-UI}
\end{figure}

The web server communicates with the user through a web browser built using various front-end technologies to display a variety of elements and controls inside a user-friendly layout. The browser is outlined by seven functional elements that guide users through navigation. The descriptions of these elements are provided in Table \ref{tab:UI-ELEMENTS}. Example views of these elements are shown in Fig.\ref{fig:A-TEAM-UI}. 

\begin{table}[H]
\caption{Descriptions of all UI elements implemented in A-TEAM}
\centering
\begin{tabular}{|p{0.35\linewidth}|p{0.65\linewidth}|}
\hline
\textbf{UI Element}         & \textbf{Description}                                                                                        \\ \hline
Mapping            & Display elements with geographical information on the map                                          \\ \hline
Search             & Search box allowing users to locate the record of events with partial ID or road name              \\ \hline
Filter             & Menu allowing users to filter events by conditions                                                 \\ \hline
Input              & Input box allowing users to type-in information                                                    \\ \hline
Edit               & Function allowing users to make changes to event information or model parameters                   \\ \hline
Spatial selection  & Function allowing users to select roadways or events by drawing a polygon or circle                \\ \hline
Temporal selection & Time bar by dragging which users can select what temporal information to display based on the time \\ \hline
\end{tabular}

\label{tab:UI-ELEMENTS}
\end{table}

\subsubsection {Visualization} 

\begin{figure}[H]
    \centering
    \includegraphics[width=1\textwidth]{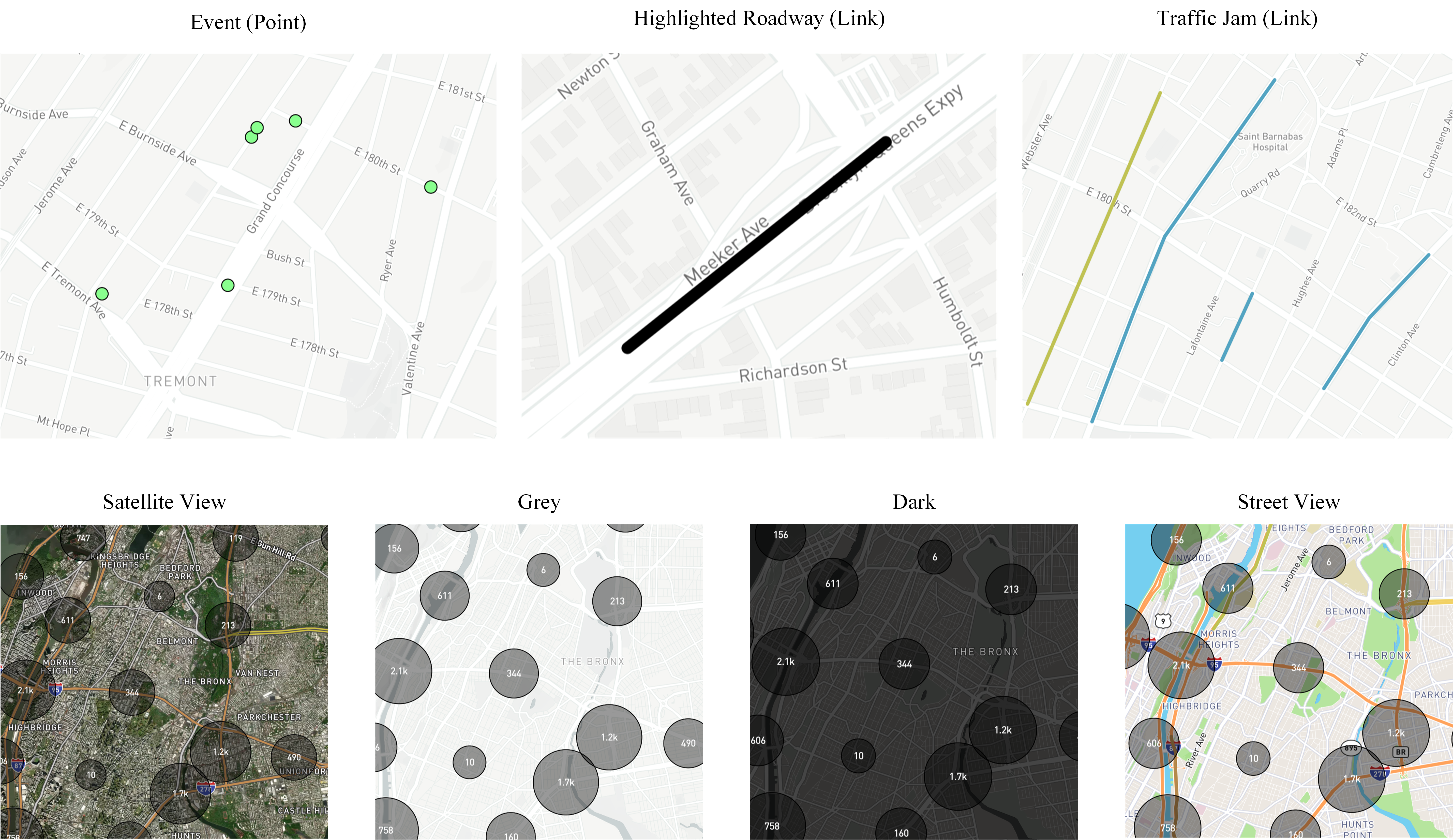}
    \caption{Examples of A-TEAM map view elements}
    \label{fig:A-TEAM-Map-View}
\end{figure}

The web server provides two types  of visualization to display data and model outputs. \emph{Map View} provides a geographical view of spatial elements on the left side of the interface. The basic map elements to be visualized include point, link, and base map. Point-based elements are mainly point-based incidents with coordinates such as work zones and accidents as shown in the top left figure of Fig.\ref{fig:A-TEAM-Map-View}. Link-based elements include links and fused link-based attributes such as travel speed, queue, and congestion status as shown in the top right two sub-figures in Fig.\ref{fig:A-TEAM-Map-View}. In addition, the web server offers four base map options, which are grey, dark, satellite, and street view as shown in the bottom four sub-figures in Fig.\ref{fig:A-TEAM-Map-View}.  

\begin{figure}[H]
    \centering
    \includegraphics[width=0.8\textwidth]{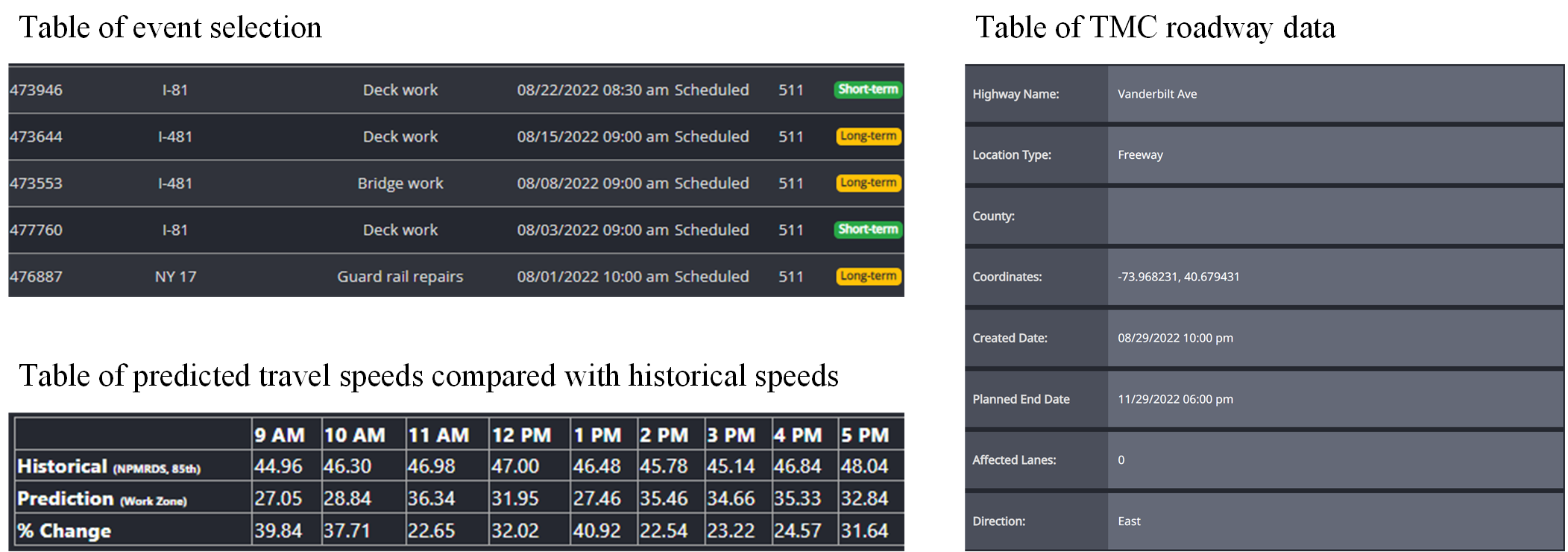}
    \caption{Examples of A-TEAM table visualizations}
    \label{fig:A-TEAM-Table-View}
\end{figure}

\begin{figure}[H]
    \centering
    \includegraphics[width=0.8\textwidth]{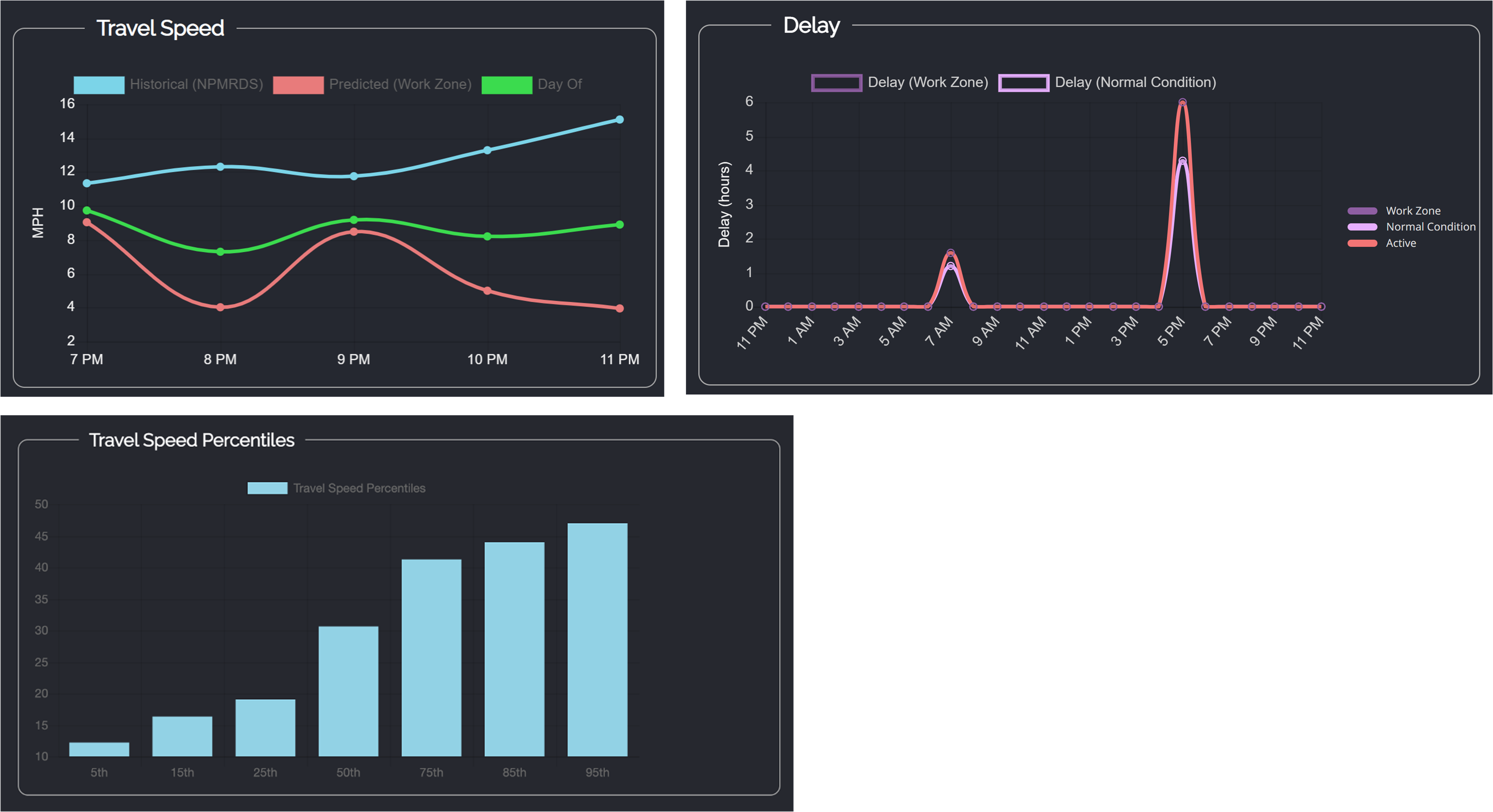}
    \caption{Examples of A-TEAM graph visualizations}
    \label{fig:A-TEAM-Graph-View}
\end{figure}

\noindent \emph{Tabular and Graphic View} visualizes data and model outputs using tables and graphs on the right side of the interface. Shown in \ref{fig:A-TEAM-Table-View}, the table elements include event information covering event type, coordinates, duration, etc., as well as roadway data which includes the name, location, length, type of the studied links, and numerical results provided by computational models. These models include predicted metrics of the impact of an event in the impact analysis tool and the percentage of overlapped impacts in the conflict analysis tool. Moreover, A-TEAM employs various types of graphs to visualize data and analysis outputs (Fig.\ref{fig:A-TEAM-Graph-View}), such as bar charts to visualize travel speed percentiles or line charts to visualize delay and travel speed over a temporal horizon which are composed of computational results returned by the model server.

\section{Case study - A work zone management and coordination application}\label{sc:case}

\begin{figure}[t]
    \centering
    \includegraphics[width=0.6\textwidth]{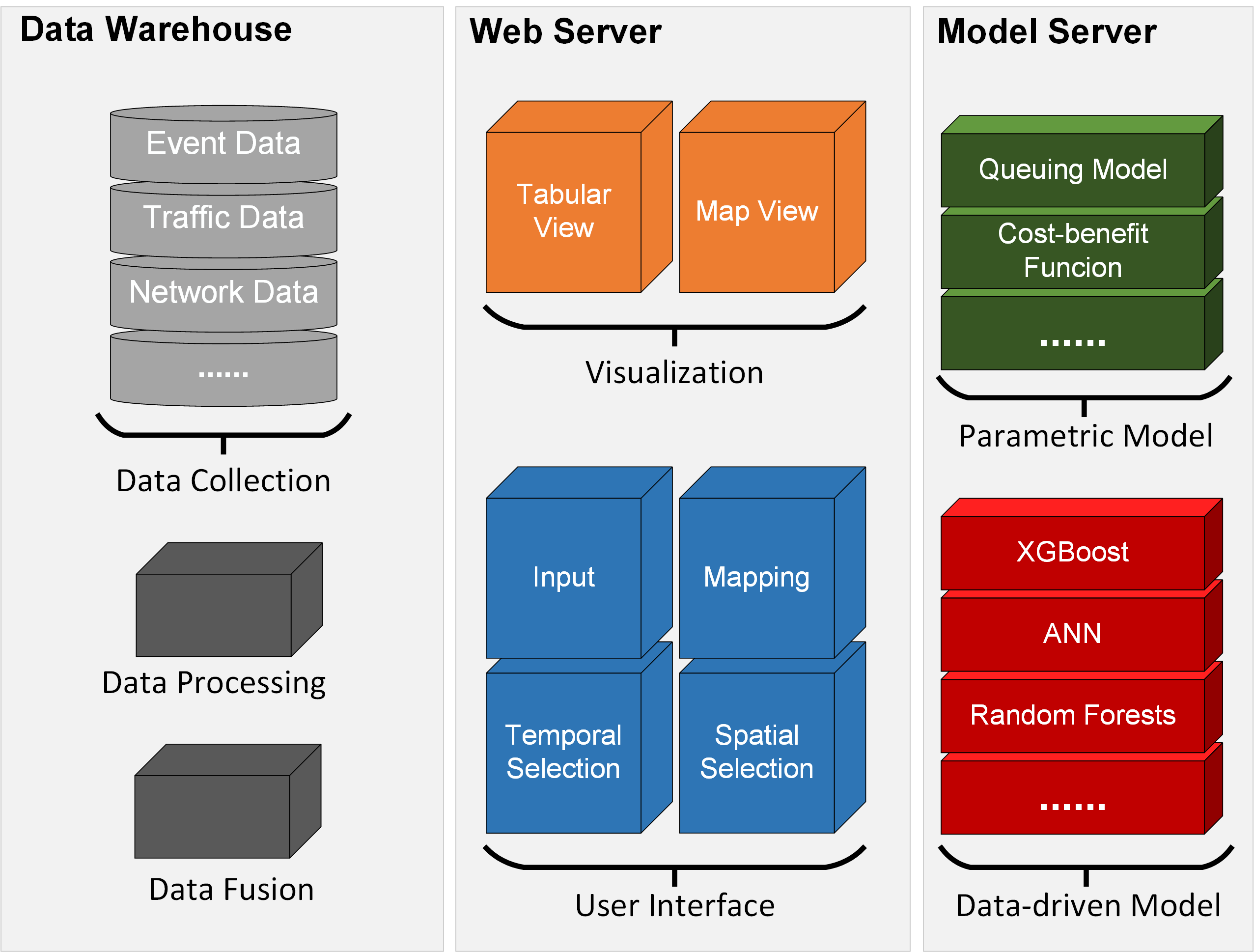}
    \caption{Application view of A-TEAM}
    \label{fig:A-TEAM-application_view}
\end{figure}

A-TEAM incorporates versatile modules to facilitate various types of numerical and graphical analysis related to the spatio-temporal analysis of traffic event impacts. Fig.\ref{fig:A-TEAM-application_view} demonstrates the application view of all sub-modules in the designed system. With the fusion of different functional modules, A-TEAM is able to produce various applications upon user needs and preferences. 

The following case study demonstrates a work zone management and coordination application that is built upon the A-TEAM platform to provide decision support to traffic engineers and traffic practitioners in analyzing and coordinating conflicted work zones. With proper deployment of the sub-modules in Fig.\ref{fig:A-TEAM-application_view}, three analysis modules have been developed for this work zone application: 1) Impact Analysis, 2) Conflict Analysis, and 3) Coordination Analysis. 

\subsection{Impact Analysis}
The incident impact analysis is designed to predict/estimate the negative impacts to roadway systems due to the occurrence of disruptive traffic events. In the work zone coordination and management application, it is utilized to quantify work zone impacts on roadway traffic, including the queue length along the roadway and the travel delay caused by the work zone. 

\begin{figure}[t]
    \centering
    \includegraphics[width=0.4\textwidth]{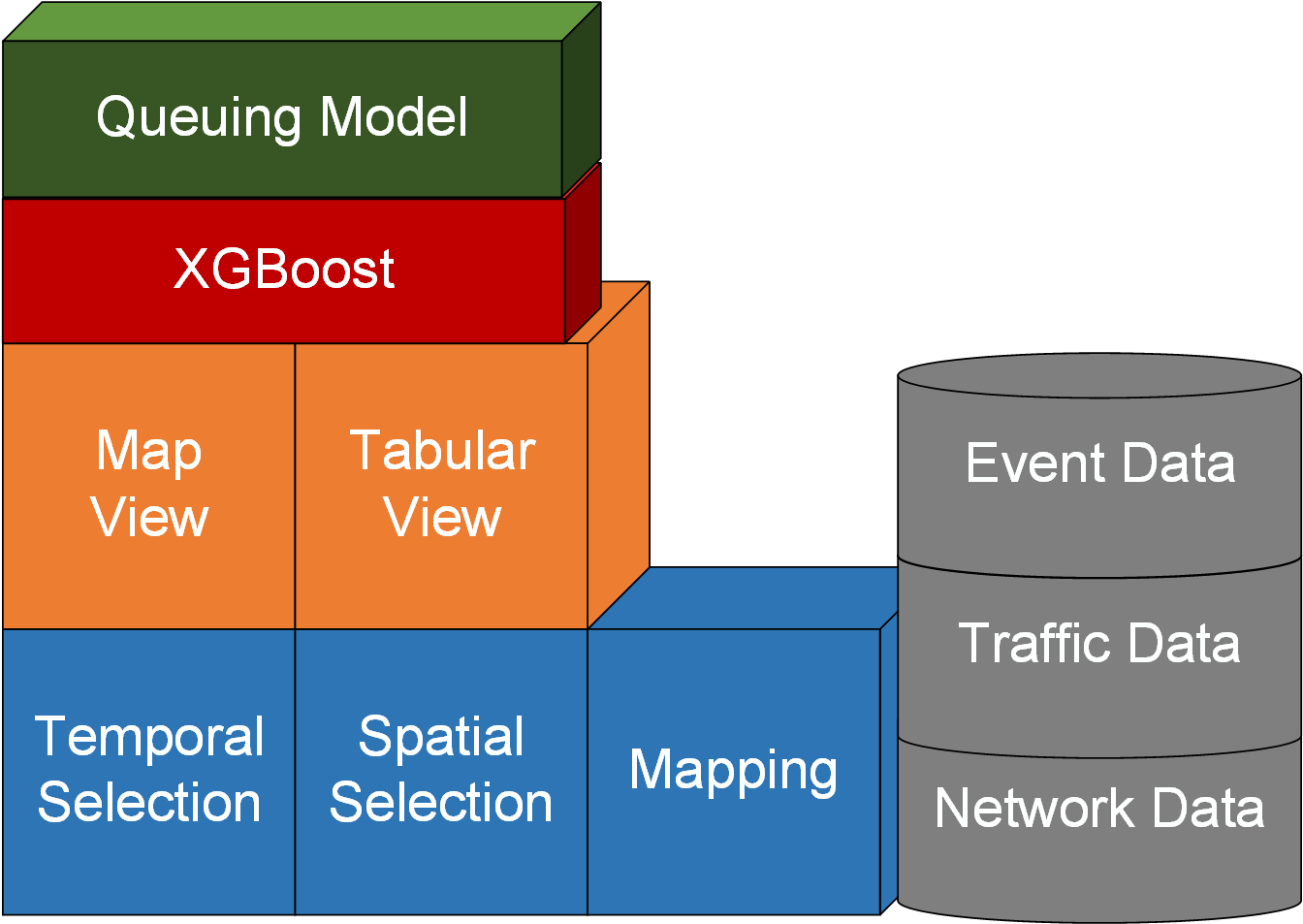}
    \caption{Application view of impact analysis}
    \label{fig:app-impact-analysis}
\end{figure}

\begin{figure}[!t]
  \centering
  \subfloat[Queuing Model]{\includegraphics[width=0.48\textwidth]{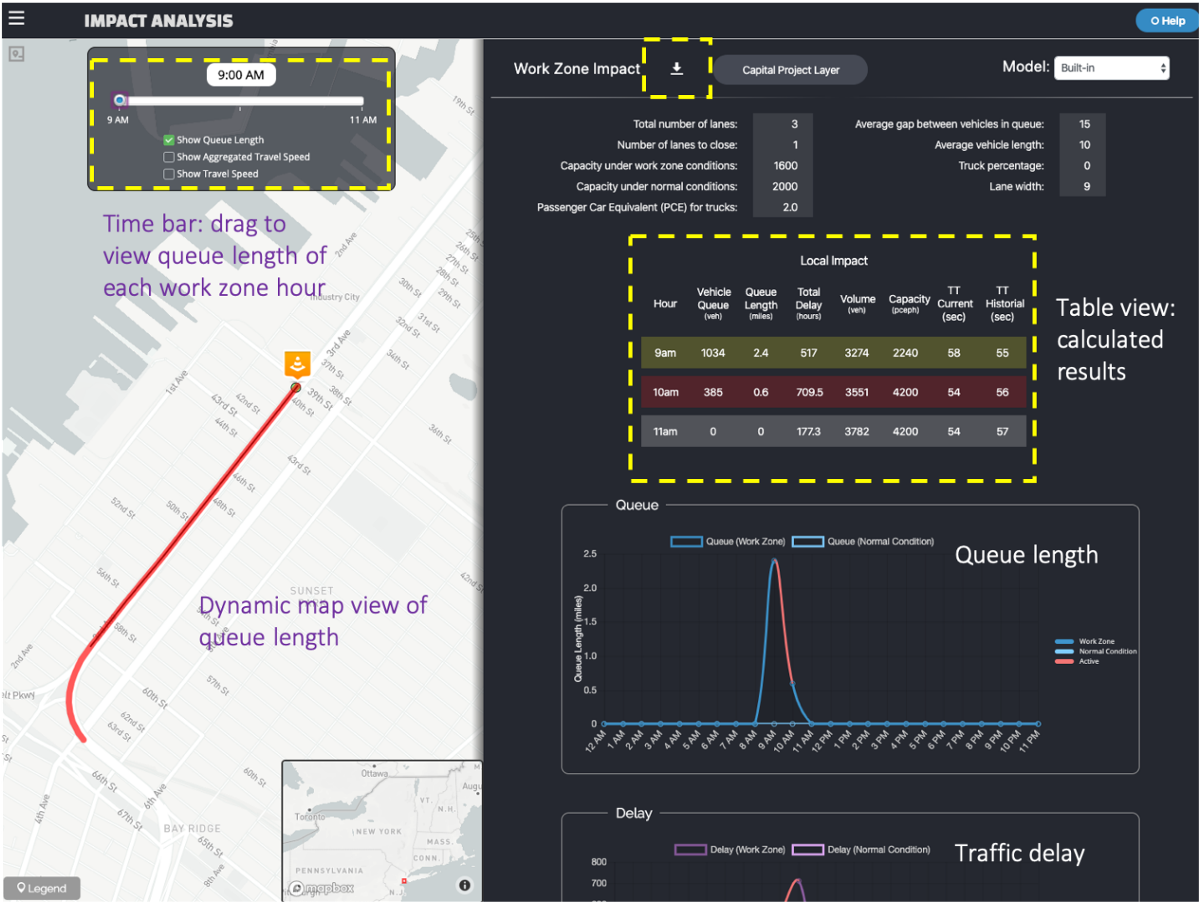}}
  \hfill
    \subfloat[XGBoost]{\includegraphics[width=0.42\textwidth]{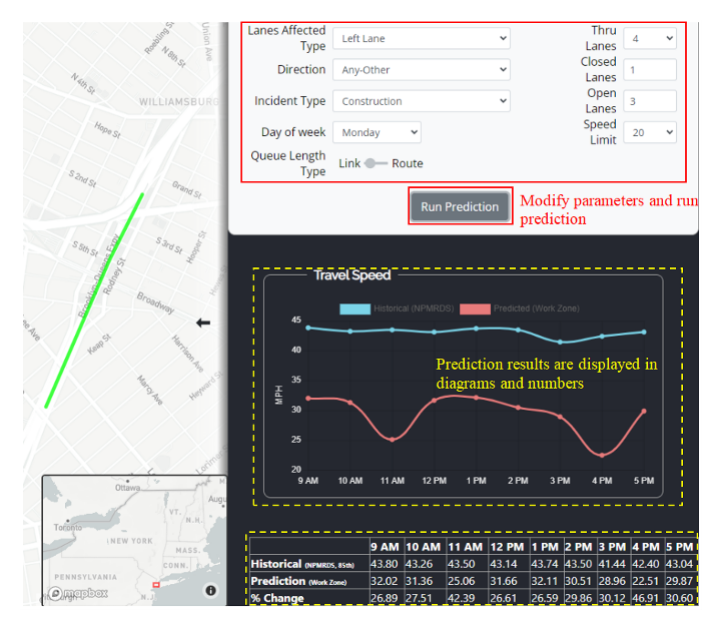}}
  \caption{Impact prediction analysis}
  \label{fig:impact-analysis}
\end{figure}

As shown in Fig.\ref{fig:app-impact-analysis}, various sub-modules in the data warehouse, web server and model server are involved to support the impact analysis. One important feature of the impact analysis module is that it can be run with and without the traffic volume information, with the power of both parametric modeling and data-driven modeling capabilities. To be specific,  when traffic volume information is available, the impact analysis function adopts the deterministic queuing models from the parametric model sub-module. Otherwise, when the traffic volume information is not available, the impact analysis employs the traffic speed information in conjunction with the available machine learning models (e.g., XGBoost, Random forest, ANN) to predict the impacts. The estimation/prediction results can then be visualized in both map view and tabular view, as shown in Fig.\ref{fig:impact-analysis}. 

Fig.\ref{fig:impact-analysis}(a) displays the traffic impact estimated by deterministic queuing model, using an example of a one-hour construction work zone on the Gowanus Expressway in Brooklyn, NY that was scheduled from 9AM to 10AM. Fig.\ref{fig:impact-analysis}(b) shows another example using XGBoost to predict traffic impacts for an eight-hour construction work zone on Brooklyn-Queens Expressway in Brooklyn, NY from 9AM to 5PM. 

\subsection{Conflict analysis}

\begin{figure}[t]
    \centering
    \includegraphics[width=0.4\textwidth]{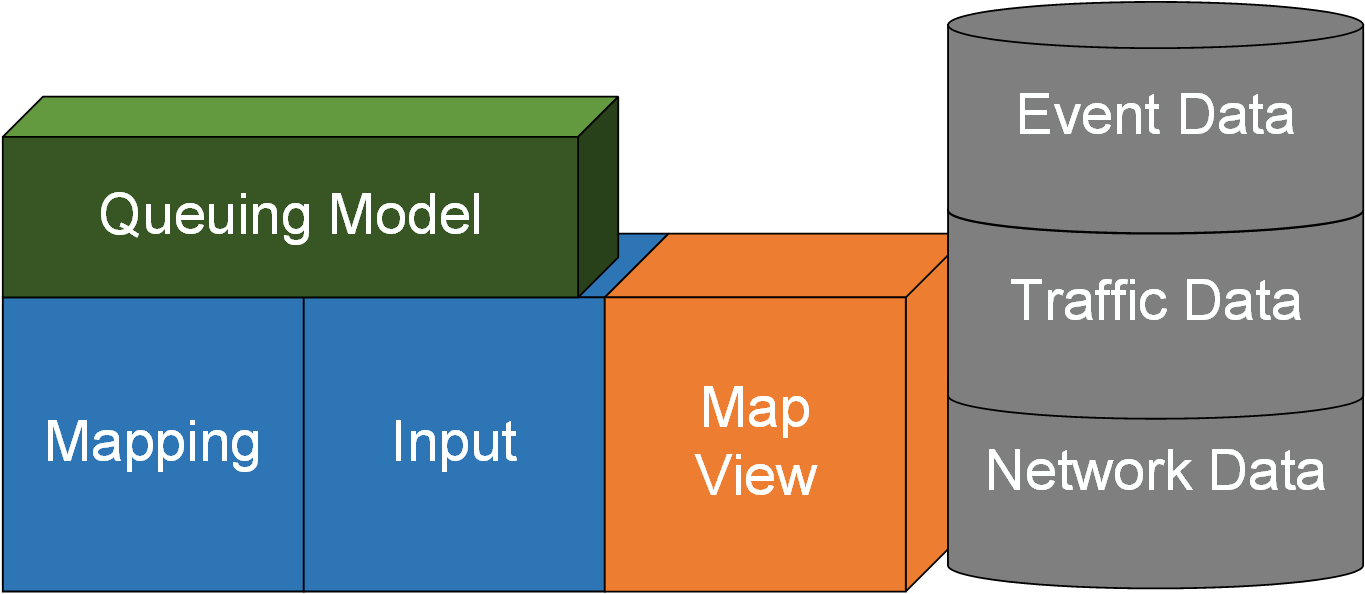}
    \caption{Application view of conflict analysis}
    \label{fig:app-conflict-analysis}
\end{figure}

The conflict identification analysis module is used to identify the potential conflicts caused by two traffic events that might overlap in both temporal and spatial dimensions. As shown in Fig.\ref{fig:app-conflict-analysis}, the conflict analysis employs different sub-modules across the data warehouse, web server and model server. Particularly, the conflict analysis extracts event data, traffic data and network / infrastructure data from the data warehouse. It then feeds the extracted data into four candidate queuing models in the model server and communicates both the input and output via the user interface and visualization functionalities implemented as part of the web server. 

\begin{figure}[!t]
  \centering
  \subfloat[]{\includegraphics[width=0.58\textwidth]{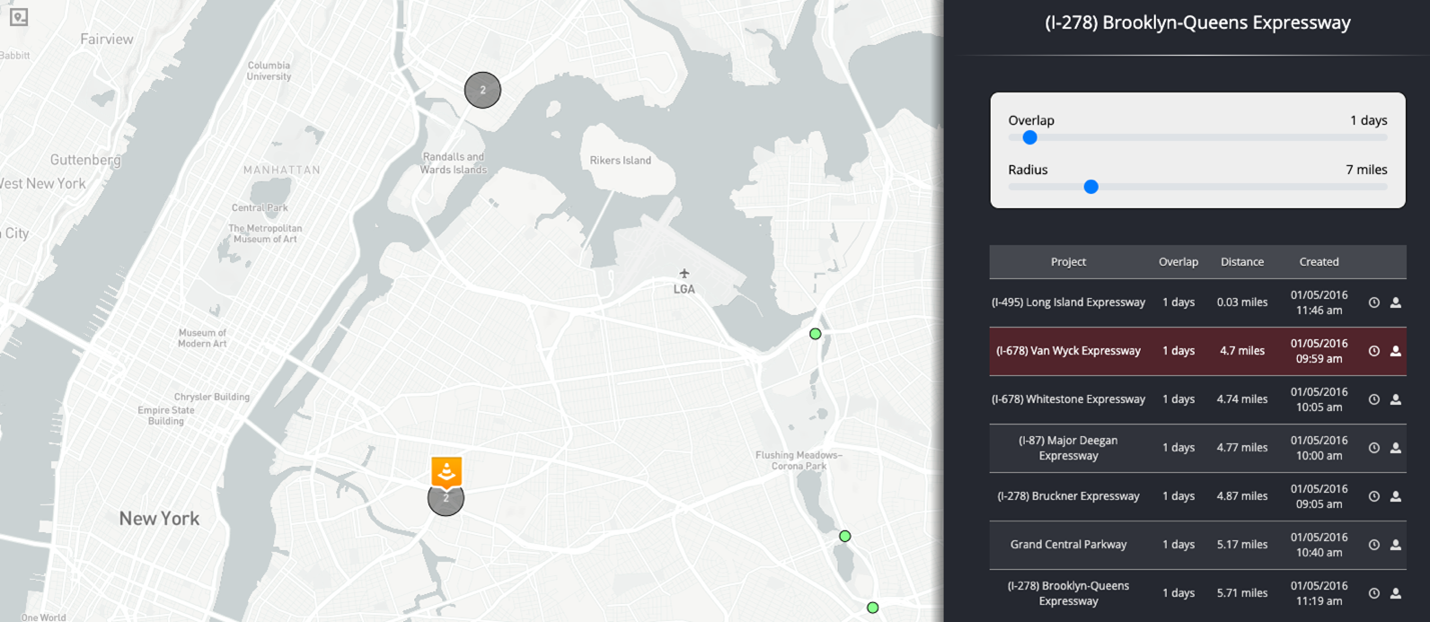}}
  \hfill
    \subfloat[]{\includegraphics[width=0.37\textwidth]{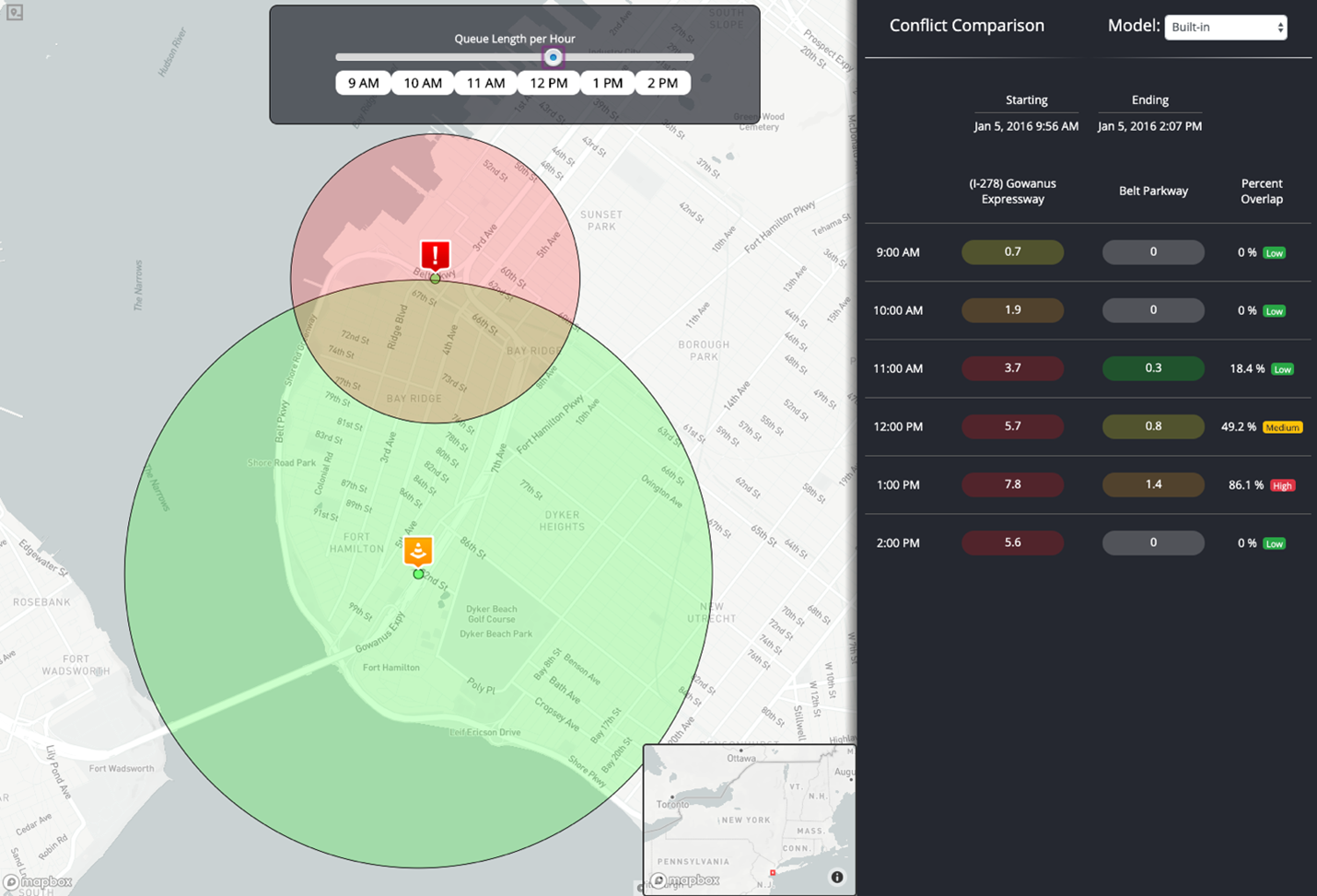}}
  \caption{Conflict identification analysis}
  \label{fig:conflict-analysis}
\end{figure}

The underlying logic of the conflict analysis is to check any overlaps in three different levels, including temporal, spatial and impact. As shown in Fig.\ref{fig:conflict-analysis} (a), given the input of time range and physical distance between the work zones, the conflict analysis identifies a total of 6 potential conflicting work zone events (shown in green dots) for a work zone event of interest. 

Furthermore, as the impact of a work zone is dynamic (e.g., queue length varies depending on the congestion level), it is not sufficient to identify potential conflicts by just investing the overlap of construction time and physical distances between work zones. We therefore initiate parametric models to estimate the potential dynamic traffic impacts expected to be caused by the concurrent existence of these work zones and check if an impact overlap is observed over time. The conflict analysis function visually depicts the area affected by the work zones: plotting circles using the estimated/predicted queue lengths caused by the conflicting traffic events and calculating the overlapped percentage between the two circles. As shown in Fig.\ref{fig:conflict-analysis}(b), the left panel displays the overlap of traffic impacts caused by two conflicting events on the map (i.e., green and red circles), the right panel displays the table view of numerical results of percentage overlap at different points in time. 

\subsection{Coordination Analysis}

The coordination analysis allows users to coordinate conflicted work zones by rescheduling or consolidating them to avoid conflicts. The aforementioned applications of conflict analysis and impact analysis are incorporated into the coordination analysis to assist the decision making process of coordination. 

\begin{figure}[H]
  \centering
  \subfloat[Conflicting work zones]{\includegraphics[width=0.65\textwidth]{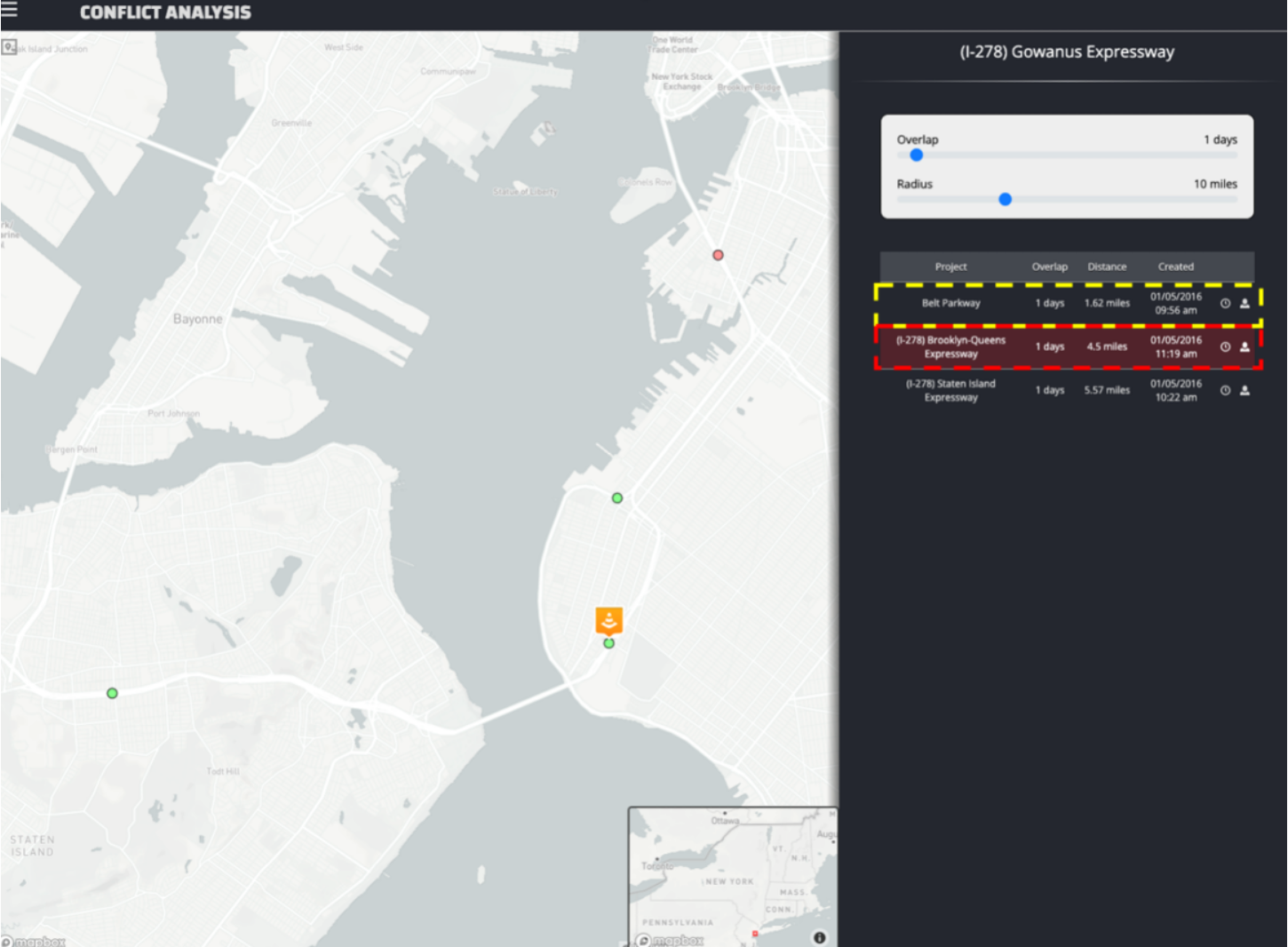}}
  \hfill
    \subfloat[Cost-benefit inputs]{\includegraphics[width=0.28\textwidth]{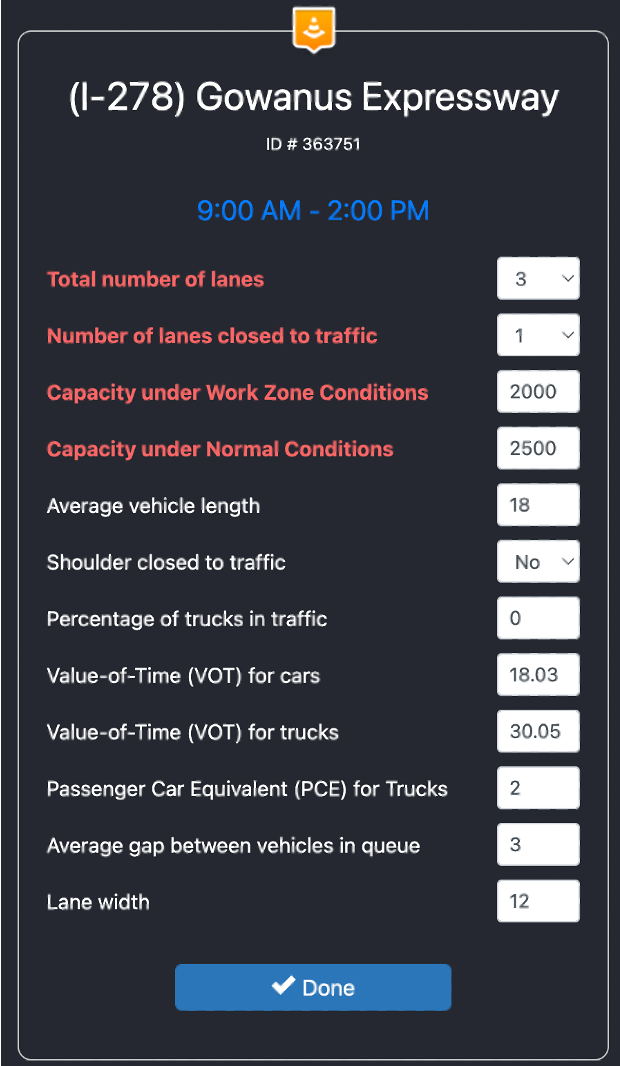}}
  \newline
  \subfloat[Before and after rescheduling]{\includegraphics[width=1\textwidth]{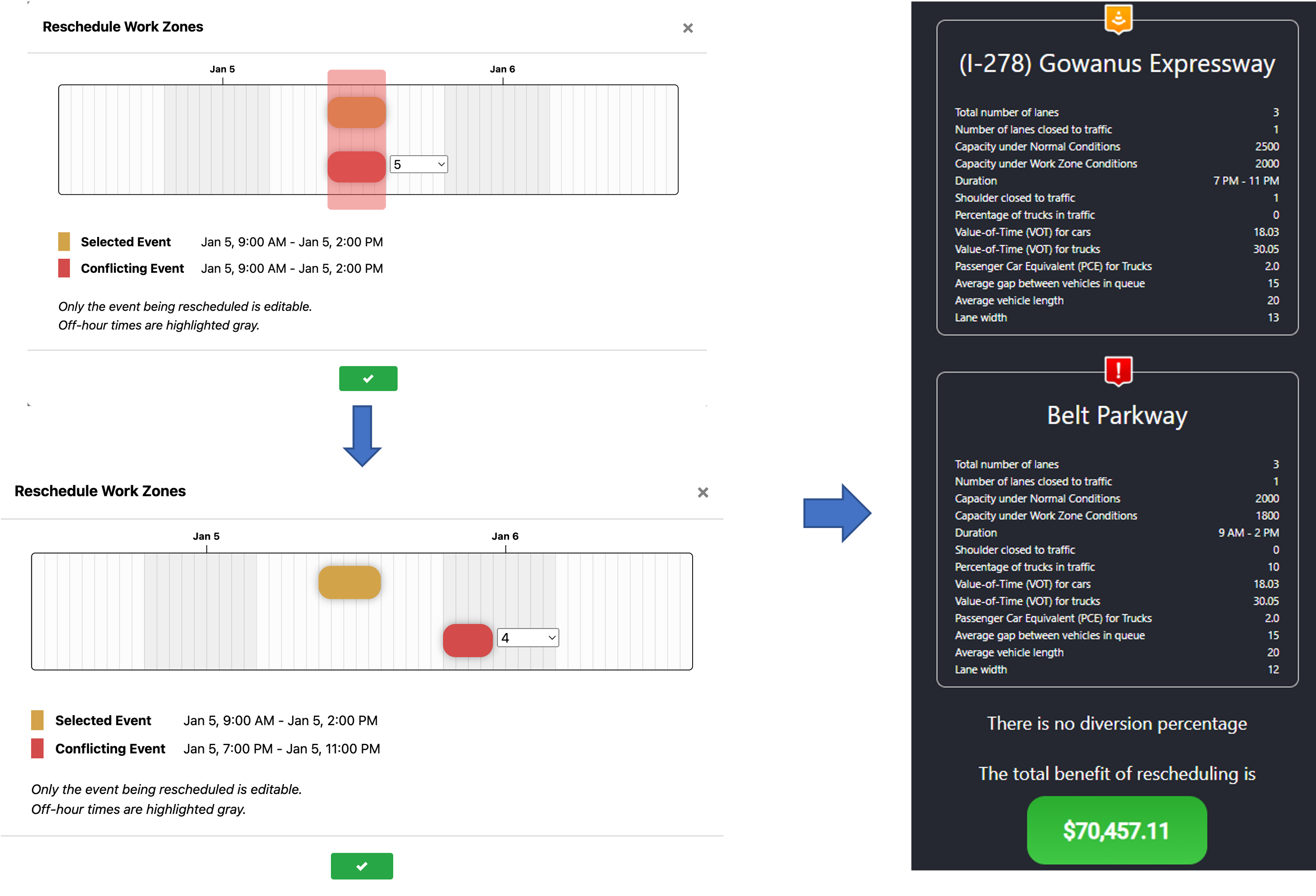}}
  \caption{Coordination analysis}
  \label{fig:coordination-analysis}
\end{figure}

As an example, we use data retrieved from \cite{bian2021cidnyreport} which include two construction work zones on Gowanus Expressway and Belt Parkway in Brooklyn, NY that have actually occurred in the past to demonstrate how they could have been coordinated if A-TEAM was used before they became active. A-TEAM first identifies a list of potential conflicting work zones that are within the distance of 2 miles and the time overlap of more than 1 hour for the work zone on Gowanus Expressway. As shown in Fig.\ref{fig:coordination-analysis} (a), a candidate work zone on Belt Parkway is identified and selected. To be specific, both work zones were active between 9AM and 2PM on January 5, 2016. The Euclidean distance between two work zones is 1.62 mile and both work zones share a five-hour time overlap. 

To quantify the extent of conflict between the two work zones, A-TEAM provides a before-after cost-benefit analysis to assist users in the decision-making process of coordinating conflicting work zones. A-TEAM first generates the total cost before and after coordinating the work zones. The total cost is calculated by assigning value of time to the delay caused by the work zones in the impact analysis module. The benefit after rescheduling the work zones to allow for better coordination is calculated as the monetary value of avoided delay. Eq.\label{e1} shows the calculation of the total cost/benefit of traffic delay caused by the work zone.

\begin{equation}\label{e1}
    TC = VOT_{car} \times TD \times (1-HV) + VOT_{HV}\times TD\times HV
\end{equation}
where $TC$ denotes the total cost of a work zone (in the unit of USD), $VOT_{car}$ is the value of time for passenger car (USD/hour), $VOT_{HV}$ is the value of time for heavy vehicles (USD/hour), $TD$ is the total delay caused by the work zone (hour), and $HV$ is the percentage of heavy vehicles.

As shown in Fig.\ref{fig:coordination-analysis}(b), A-TEAM has an interactive user interface allowing users to reschedule the conflicting work zones by dragging the time bar (shown in yellow and red). The left of Fig.\ref{fig:coordination-analysis}(c) shows the before and after of rescheduling the conflicting work zone. The semitransparent red band that covers both yellow and red bars in the top left of Fig.\ref{fig:coordination-analysis}(c) represents the time overlap of these two work zones. After rescheduling the conflicting work zone, as shown in the bottom left of Fig.\ref{fig:coordination-analysis}(c), the semitransparent red band disappeared since the red bar was dragged to a new schedule at 7PM-11PM. A-TEAM reported a total benefit of rescheduling the conflicting work zone as \$70,457.11, as shown in the right part of Fig.\ref{fig:coordination-analysis}(c). The result of the cost-benefit analysis indicates that moving the schedule of a work zone to the evening off peak period can avoid the daytime congestion and thereby produces less negative impact to traffic and less conflict with surrounding work zones as compared to using the daytime work zone schedule.

Using the interactive user interface to adjust the scheduling information of the conflicting work zones, users can easily  test and try different combinations  of work zone schedules. In this way, A-TEAM can help users to better understand the consequence of coordination and provide quantitative support for their decision making process.

\section{Potential capabilities}
The modular design of A-TEAM provides flexibility and scalability to customize the applications for different user needs and preferences, and various potential traffic management applications can be easily built upon it. We will introduce how two potential applications, a network-wide traffic forecasting and a traffic accident detection and prediction application, can be built using the A-TEAM platform.

\subsection{Network-wide traffic forecasting}
A-TEAM has the potential of incorporating the application of forecasting traffic mobility information for a defined transportation network. The required data includes the traffic mobility data (e.g., traffic speed, travel time or traffic volume) and spatial data (e.g., geometric information and network topology). As mentioned in the previous sections, the traffic data in the current data warehouse already provides available link-based traffic speed data, link geometry, and network topology information. The GPU and CPU clusters in the model server provide the computational power for executing user-customized AI/ML data-driven models. The web server provides a pre-defined user interface allowing users to interact and visualize the prediction results.

\begin{figure}[H]
  \centering
  \hspace*{\fill}
  \subfloat[Application view]{\includegraphics[width=0.3\textwidth]{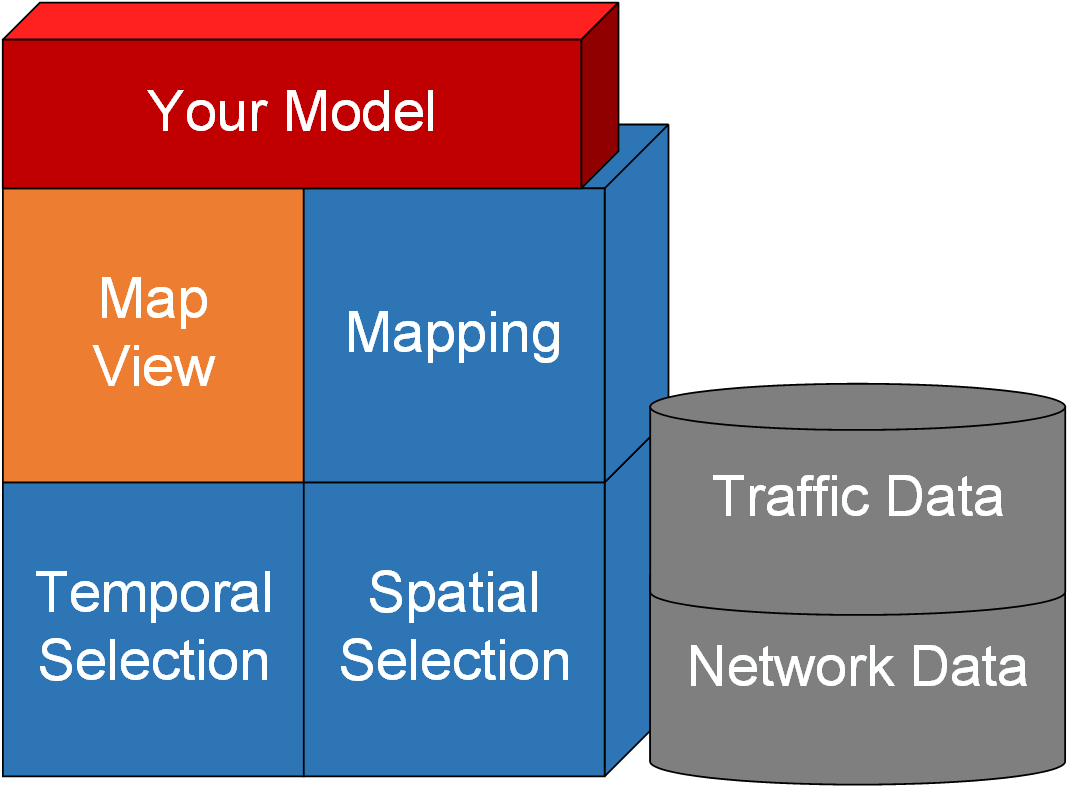}}
  \hfill
    \subfloat[Example visualization]{\includegraphics[width=0.3\textwidth]{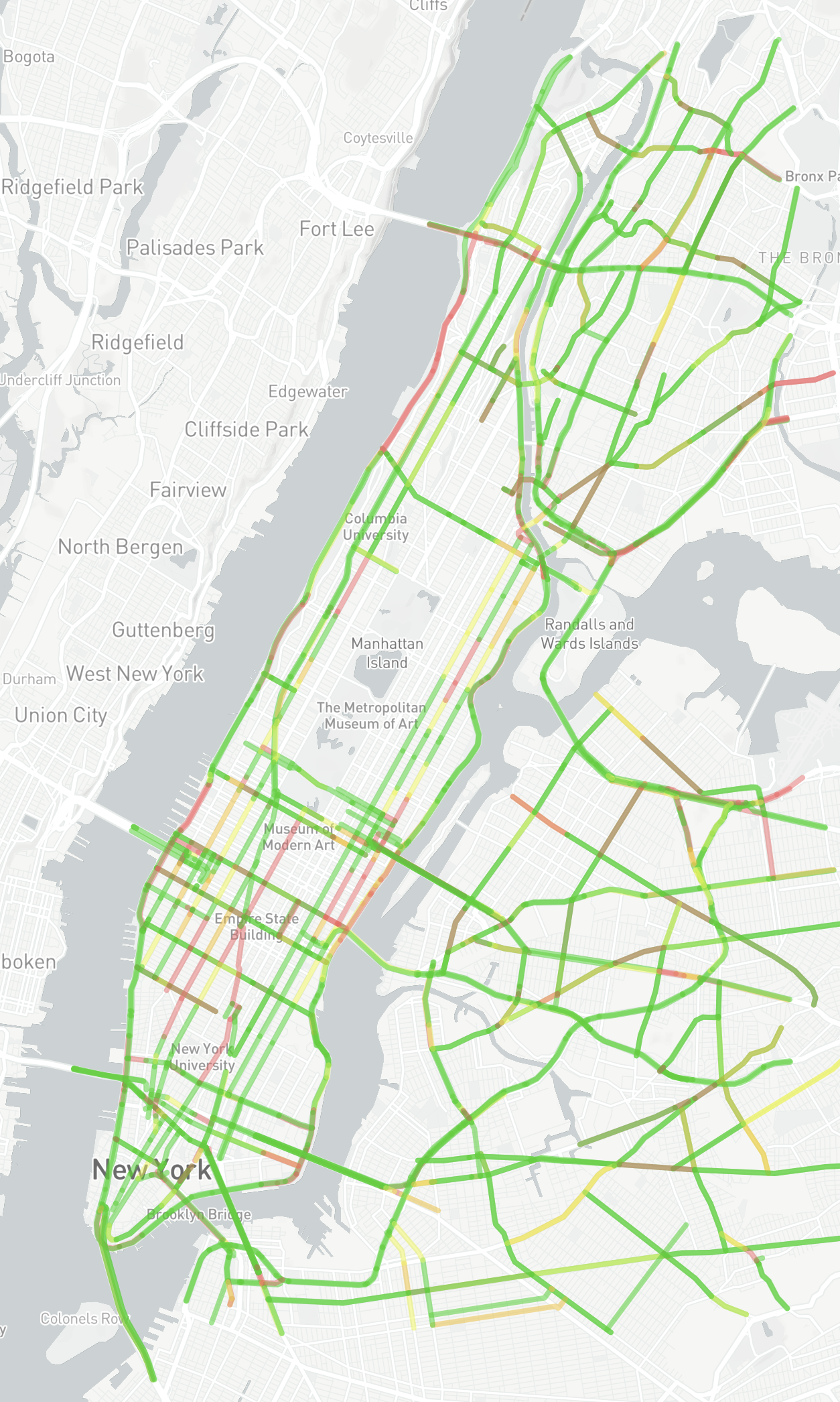}}
\hspace*{\fill}
  \caption{Potential application 1: network-wide traffic forecasting}
  \label{fig:potential-application1}
\end{figure}

Therefore, once the users provide their modeling file, for example, a python file stored pre-trained AI/ML model parameters, A-TEAM is able to implement and compile the modeling file with other existing modules and compose a traffic forecasting application. Fig.\ref{fig:potential-application1}(a) illustrates the proposed application view of the required modules and elements for traffic forecasting and Fig.\ref{fig:potential-application1}(b) presents an example visualization of predicted traffic speeds around Manhattan area in New York City.

\subsection{Traffic accident detection and prediction}
\begin{figure}[H]
  \centering
  \hspace*{\fill}%
  \subfloat[Application view]{\includegraphics[width=0.34\textwidth]{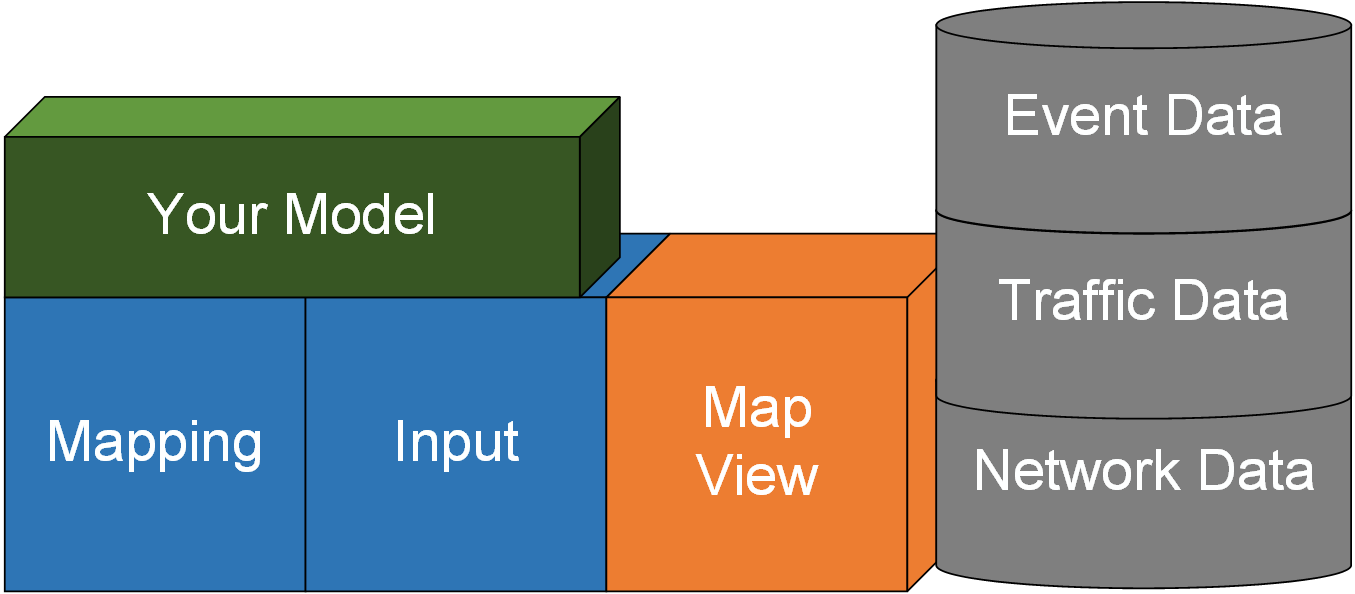}}
  \hfill
  \subfloat[Example visualization]{\includegraphics[width=0.3\textwidth]{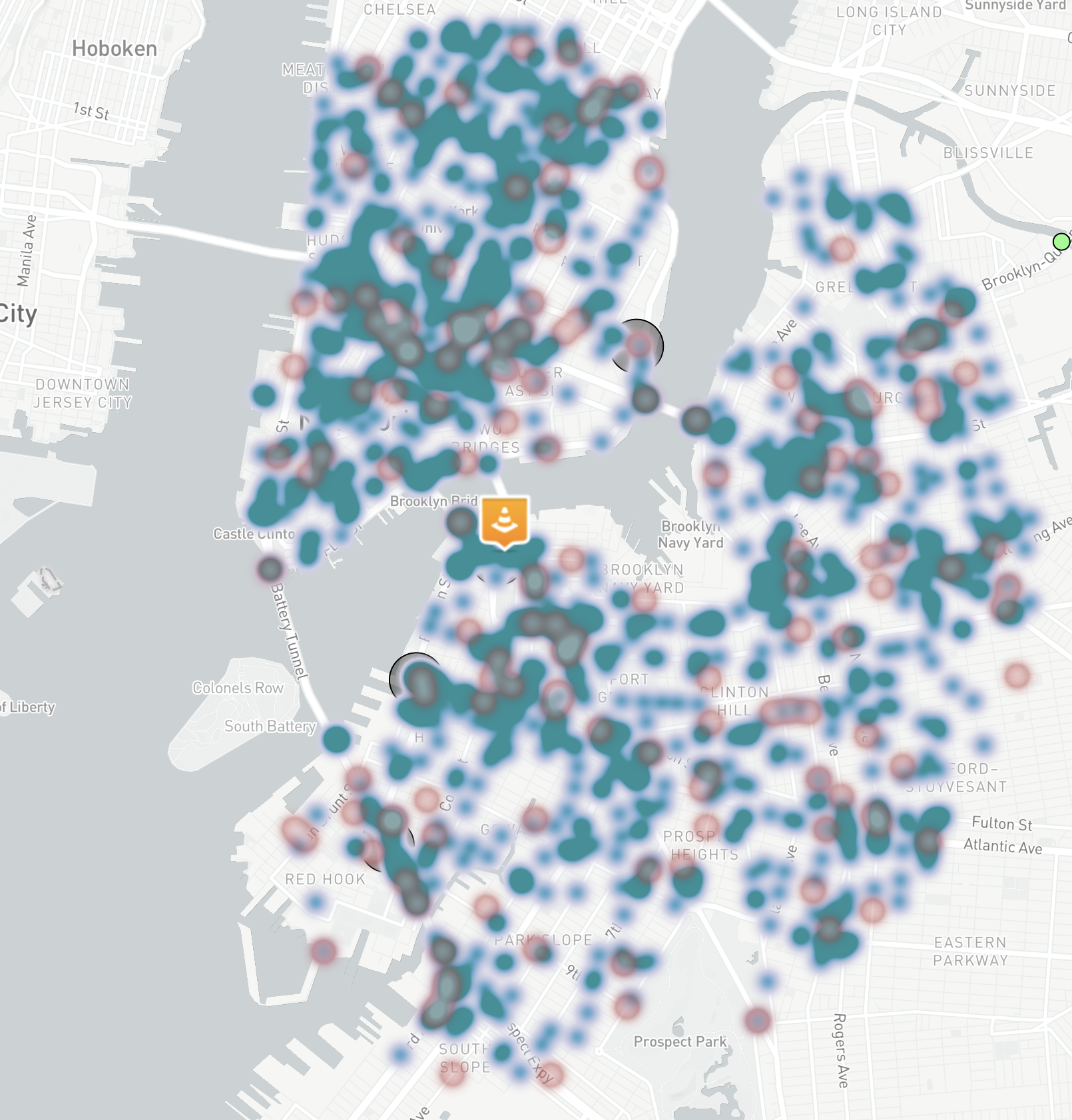}}
  \hspace*{\fill}%
  \caption{Potential application 2: traffic accident detection and prediction}
  \label{fig:potential-application2}
\end{figure}

A second potential application that can be provided by A-TEAM is traffic accident detection and prediction. As shown in Fig.\ref{fig:potential-application2}(a), the current event data in the data warehouse of A-TEAM already provides traffic crash information. If users provide their crash detection and prediction modeling files, A-TEAM is able to implement the pre-trained file and visualize the detected/predicted results directly.

To be specific, A-TEAM's user interface in the web server allows users to perform both temporal and spatial selection of traffic events, including traffic accidents. Fig.\ref{fig:potential-application2}(b) presents an example visualization of detected/predicted results of crashes, if given the specific prediction temporal horizon and spatial neighborhood. The blue dots can be used to represent the detected crashes and the red dots can be used for predicted crashes. Therefore, A-TEAM is user-friendly to incorporate user-provided modeling files and implement them into applicable usage.

\section{Conclusions} \label{sc:conclusion}
The era of 'big data' expedites the development of transportation-related solutions, specifically in the form of data-driven models with high performance in terms of accuracy, flexibility and scalability. However, the deployment of such data-driven models in transportation applications usually requires the use of strict input data formats and programming knowledge, making them difficult to be deployed in the real-world. In other words, a significant gap exists between the state-of-the-art and state-of-the-practice. The traffic researchers have been constructing and validating data-driven models for a long time. However, deployment of these models for real-world application may have been significantly delayed, or never realized, due to high time efforts and budget requirements when it comes to the development of robust and stable software tools by the research community. On the other hand, traffic engineering practitioners who are familiar with the task of developing traffic engineering tools for real-world applications may lack the domain knowledge of developing advanced data-driven AI/ML models that have been becoming more sophisticated and powerful during the last decade.

To bridge this gap, this paper proposes a modular platform called A-TEAM which aims to accelerate the path from research prototyping to real-world application deployment and can be customized by both traffic researchers and practitioners based on their needs and preferences. The model server in A-TEAM is equipped with high performance computing resources, supporting the incorporation and subsequent execution of the pre-trained model files into the integrated A-TEAM platform. A robust data-model pipeline connects the pre-trained model files with the data warehouse in the back end and the web user interface in the front end. In this way, A-TEAM platform aims to reduce the workload for traffic researchers and traffic practitioners while deploying their data-driven models into a functional transportation software. For illustration purposes, this paper employed a comprehensive work zone coordination and management application which is built on the A-TEAM platform. This application helps users in identifying conflicting work zones, estimating traffic impacts caused by the work zones and coordinating conflicting work zones.  

Currently, A-TEAM only incorporates the comprehensive analysis and application of coordinating work zones, but this paper also discusses two potential applications namely, network-wide traffic forecasting and traffic accident detection and prediction application, that can be easily built on the A-TEAM platform if corresponding model files are provided by the users. The future work of A-TEAM is to keep implementing multiple transportation applications by adding more types of traffic data and state-of-art data-driven models. In addition, the design of the web user interface can also be improved to support additional analysis.

\section*{Acknowledgements}
This study was partially supported by the University Transportation Research Center (UTRC) at the City University of New York, CIDNY (Coordinated Intelligent Transportation Systems Deployment in New York City) program, and C2SMART, a Tier 1 University Transportation Center at New York University. The contents of this paper only reflect views of the authors who are responsible for the facts and do not represent any official views of any sponsoring organizations or agencies. \par

\section*{Author contributions}
The authors confirm contribution to the paper as follows: \textbf{study conception and design}: Zilin Bian, Kaan Ozbay; \textbf{data curation}: Zilin Bian, Dachuan Zuo, Matthew D. Maggio; \textbf{analysis and interpretation of results}: Zilin Bian, Dachuan Zuo, Jingqin Gao; \textbf{draft manuscript preparation}: Zilin Bian, Dachuan Zuo, Jingqin Gao, Kaan Ozbay. All authors reviewed the results and approved the final version of the manuscript.

\newpage

\bibliographystyle{trb}
\bibliography{main}
\end{document}